\documentclass[10pt,a4paper]{article}
\usepackage{times}
\usepackage{graphicx}
\usepackage{amsmath}
\usepackage{amssymb, bm}
\usepackage{mathrsfs}
\usepackage{booktabs}
\usepackage{array}
\usepackage{multirow}
\usepackage{lineno}
\usepackage{float}
\usepackage{pdflscape}
\usepackage[numbers,sort&compress]{natbib}
\usepackage{setspace}
\usepackage[colorlinks=true, allcolors=blue]{hyperref}
\usepackage{titlesec}
\selectfont

\usepackage{titlesec}
\usepackage{etoolbox}
\usepackage{setspace}
\usepackage[left=2.cm,right=2.cm,top=2.cm,bottom=2.cm]{geometry}

\titleformat{\section}
{\fontsize{10}{9}\selectfont\bfseries\raggedright}
{\thesection.}{0.3em}{}
\titlespacing*{\section}{0pt}{10pt}{9pt}

\titleformat{\subsection}
{\fontsize{10}{9}\selectfont\bfseries\raggedright}
{\thesubsection.}{0.3em}{}
\titlespacing*{\subsection}{0pt}{10pt}{9pt}

\titleformat{\subsubsection}
{\fontsize{10}{9}\selectfont\bfseries\itshape\raggedright}
{\thesubsubsection.}{0.3em}{}
\titlespacing*{\subsubsection}{0pt}{10pt}{6pt}

\title{Differential coupling drives Turing pattern formation in neural fields: a mechanism for Parkinson's disease and deep brain stimulation}
\author{
	Ju Kang\textsuperscript{1,*}, 
    Anmin Zheng\textsuperscript{1} 
}
\date{}

\newcommand{\affil}[2]{%
	\textsuperscript{#1}#2%
}

\makeatletter
\renewcommand{\maketitle}{
	\begin{center}
		{\LARGE\bfseries \@title\par}
		\vspace{1em}
		{\normalsize \@author\par}
		\vspace{-0.3em}
		\setlength{\parskip}{1pt}
		\@date
	\end{center}
}
\makeatother

\begin{document}
	
	\maketitle
	
	\begin{center}
		\affil{1}{Department of Mathematics and Physics, School of Artificial Intelligent Medicine, Guilin Medical University, Guilin 541199, China}\\
		%\affil{2}{School of Physics, Sun Yat-sen University, Guangzhou 510275, China}\\
		\affil{*}{Corresponding author: \href{mailto:112026040@glmu.edu.cn}{112026040@glmu.edu.cn}}\\
	\end{center}
	
	\begin{abstract}
	Deep brain stimulation (DBS) is an effective treatment for the motor symptoms of Parkinson's disease (PD), yet its mechanism of action remains incompletely understood. Abnormal beta-band synchronization in the basal ganglia and thalamocortical circuit is a hallmark of PD, but a theoretical framework linking network synchronization to spatial pattern formation is lacking. Here we show that a two-layer neural field model with differential interlayer coupling undergoes a Turing instability, spontaneously generating a diverse repertoire of spatially ordered patterns, including stripes, spots, rings, and spiral waves. The model yields the critical conditions and wavelengths for this instability, quantitatively predicting the observed pattern morphologies. Crucially, the interlayer difference, a natural measure of synchrony, exhibits strongly regime dependent dynamics. These findings establish a dynamical framework linking spatial pattern formation to pathological beta-band synchronization, providing a theoretical mechanism for DBS modulation of network states and a mathematical basis for optimizing stimulation parameters.
	\vspace{0.3cm}
	\end{abstract}
	
	\section{Introduction}
	
	Parkinson's disease is characterized by the loss of dopaminergic neurons and by a cascade of network-level changes in the basal ganglia–thalamocortical circuit. Among these changes, exaggerated beta-band (13-30 Hz) synchronization is one of the most robust electrophysiological signatures~\cite{brown2007}. Beta power is elevated in the subthalamic nucleus (STN) and in connected cortical regions of patients withdrawn from medication, and it is reduced by both dopaminergic therapy and effective deep brain stimulation (DBS) in parallel with clinical improvement~\cite{kuhn2008}. The pathological significance of beta activity lies not in its presence per se but in its temporal and spatial organization: beta occurs in bursts whose duration and inter-burst intervals are altered in the parkinsonian state, and the bursts are coherent across distributed nodes of the cortico--basal ganglia network~\cite{tinkhauser2018}. This spatial coherence is the feature that distinguishes pathological beta from physiological beta and that most directly suggests a network-level, rather than a single-nucleus, description of the disease.
	
	The cellular and molecular drivers that give rise to this network-level dysfunction are increasingly understood to involve non-neuronal mechanisms. Microglial NUS1 deficiency drives dopaminergic neurodegeneration through intercellular iron dysregulation, specifically via the LCN2-SLC22A17 axis, leading to neuronal ferroptosis~\cite{nus1_2026}. Spatial multi-omics studies have identified disease-associated oligodendrocytes that expand with pathology, where IL-33--ST2 signaling shapes $\alpha$-synuclein clearance and metabolic dysregulation in the substantia nigra~\cite{oligo_2026}. Chronic demyelination induces a distinct microglial state characterized by lysosomal dysfunction and lipid sequestration, which is reversible by pharmacological intervention~\cite{demyel_2026}. These findings collectively indicate that PD pathology involves a spatially organized, multi-cellular response that unfolds across distinct brain regions and cell types, rather than a purely neuronal process.
	
	DBS is the most effective surgical treatment for the motor symptoms of Parkinson's disease, yet its mechanism remains debated. A body of evidence indicates that clinically effective STN stimulation does not simply silence neural activity; it shifts STN firing from rhythmic synchrony to irregular, asynchronous discharge and acutely reduces beta power in the STN and motor cortex~\cite{kuhn2008}. The therapeutic rationale is therefore commonly framed as desynchronization: the disruption of an excessive, self-stabilizing synchronous state rather than the suppression of activity as such~\cite{tass2003}. What this framing leaves unspecified is the spatial dimension. Beta synchrony is not uniform across the circuit; it is organized into coherent ensembles with a characteristic spatial extent~\cite{tinkhauser2018}, and the response of this spatial organization to parameter changes (such as those produced by DBS) is not captured by purely temporal or single-node measures of synchrony.
	
	Neural field theory provides a natural language for relating temporal synchrony to spatial structure. In the Wilson-Cowan formulation, a population of neurons is described by a mean firing rate whose dynamics depend on the total synaptic input through a nonlinear gain function~\cite{wilson1972}. This mean-field reduction paradigm has been highly successful in computational neuroscience, with reduced two-variable neural mass models capturing complex cognitive processes such as perceptual decision-making~\cite{wong2006}. A nonlocal extension of this formulation admits Turing-type instabilities~\cite{turing1952} in which a spatially homogeneous state becomes unstable to a patterned state through the competition between short-range excitation and longer-range inhibition~\cite{hutt2005,cross1993}. Recent work has further shown that Turing patterns can emerge in non-traditional topologies, including simplicial complexes where links connect more than two nodes simultaneously~\cite{perc2023turing}, and in two-layer neuronal networks where field coupling induces pattern formation across layers~\cite{perc2018fieldcoupling}. This mechanism has been used to interpret a range of cortical phenomena, from visual hallucinations to epileptic dynamics, and it has been suggested that interactions between Turing and Hopf instabilities may underlie pathologies characterized by abnormal levels of neural synchrony, including Parkinson's disease~\cite{tass2003}. The two-layer architecture of the basal ganglia-thalamocortical loop, however, introduces a feature that single-layer Turing models do not address: the interaction between layers that compare their activities, so that the interlayer coupling is proportional to the difference between the two activities rather than to either activity alone.
	
	Here we develop a two-layer neural field model with differential interlayer coupling and derive its linear stability conditions. We show that the differential coupling enters the dispersion relation through renormalized intra-layer couplings, so that the Turing condition is met when the differential cross-terms overcome the stabilizing intra-layer terms. For parameters that satisfy this condition, the model generates a repertoire of spatially structured patterns, including stripes, spots, rings, radial modes, spiral waves, and branched structures, whose selection is controlled by the sigmoid nonlinearity and by the competition between Turing and Hopf instabilities. The stripe width is predicted analytically from the critical wavenumber and agrees quantitatively with numerical measurements. The interlayer difference provides a natural order parameter for interlayer synchrony, and its response to parameter changes is regime-dependent: increasing the same control parameter can either enhance or weaken synchrony depending on which pattern morphology is selected. We use this regime dependence to interpret the variable relationship between beta suppression and clinical improvement in DBS~\cite{kuhn2008}, and to suggest that the spatial morphology of residual activity, rather than beta power alone, is a more informative target for stimulation design.

	\section{Results}
	\subsection{Model framework}		
	We consider the Wilson-Cowan~\cite{wilson1972} description of a local excitatory-inhibitory population,
	\begin{equation}\label{eq:WC}
		\tau\,\frac{du(t)}{dt}  = -u(t) + f\!\left(w u(t) + I(t)\right),
	\end{equation}
	where \(u(t)\) is the mean population rate, \(\tau\) the population time constant, \(w\) the effective synaptic weight, \(f(\cdot)\) a sigmoidal input-output function, and \(I(t)\) the external drive. The term $-u$ represents the passive relaxation back to the resting state; the argument of \(f\) is the total synaptic input, which cannot in general be separated into a recurrent and an external contribution.	
	
	To account for spatial structure, we let the local activity depend on position \(\mathbf{r}\in\Omega\subseteq\mathbb{R}^{d}\) (\(d=1,2,3\)). Neurons at \(\mathbf{r}\) receive inputs from all other locations through a connectivity kernel \(w(\mathbf{r}-\mathbf{r}')\), which is assumed to be translation invariant and to decay with distance. Equation~\eqref{eq:WC} generalises to the integro-differential neural field
	\begin{equation}\label{eq:NF}
		\tau\,\frac{\partial u(\mathbf{r},t)}{\partial t} = -u(\mathbf{r},t) + f\!\left(\int_{\Omega} w(\mathbf{r}-\mathbf{r}')\,u(\mathbf{r}',t)\,d\mathbf{r}' + I(\mathbf{r},t)\right).
	\end{equation}
	The convolution in Eq.~\eqref{eq:NF} encodes the nonlocal character of cortical and subcortical wiring  (see SM Sec.~\ref{sec:expansion}).

	For a symmetric kernel, \(w(\boldsymbol{\xi})=w(-\boldsymbol{\xi})\) with \(\boldsymbol{\xi}=\mathbf{r}-\mathbf{r}'\), which decays rapidly with \(|\boldsymbol{\xi}|\), the convolution admits a gradient expansion. Expanding \(u(\mathbf{r}-\boldsymbol{\xi},t)\) about \(\mathbf{r}\) and retaining terms up to second order,
	\begin{equation}\label{eq:expansion}
		\int_{\Omega} w(\boldsymbol{\xi})\,u(\mathbf{r}-\boldsymbol{\xi},t)\,d\boldsymbol{\xi}
		= W_0\,u(\mathbf{r},t) + D\,\nabla^2 u(\mathbf{r},t) + \mathcal{O}\!\left(|\boldsymbol{\xi}|^{4}\right),
	\end{equation}
	where the zeroth and second moments are
	\begin{equation}\label{eq:moments}
		W_0 = \int w(\boldsymbol{\xi})\,d\boldsymbol{\xi}, \qquad
		D = \frac{1}{2d}\int |\boldsymbol{\xi}|^{2}\,w(\boldsymbol{\xi})\,d\boldsymbol{\xi}.
	\end{equation}
	The first-order term vanishes because \(w\) is even. For an isotropic kernel, \(D\) reduces to \(D=W_2/(2d)\) with \(W_2=\int|\boldsymbol{\xi}|^{2}w\,d\boldsymbol{\xi}\), and \(D>0\) whenever \(W_2>0\). The local approximation~\eqref{eq:expansion} is valid when the kernel width is much smaller than the spatial scale of interest; higher-order terms are then negligible.
	
	Substituting Eq.~\eqref{eq:expansion} into Eq.~\eqref{eq:NF} yields the reaction-diffusion equation
	\begin{equation}\label{eq:RD}
		\tau\,\frac{\partial u(\mathbf{r},t)}{\partial t} = -u(\mathbf{r},t) + f\!\left(W_0\,u(\mathbf{r},t) + D\,\nabla^2 u(\mathbf{r},t) + I(\mathbf{r},t)\right).
	\end{equation}
	The Laplacian term in Eq.~\eqref{eq:RD} is therefore not introduced by hand: it follows directly from the second-moment approximation of the nonlocal kernel. In Eq.~\eqref{eq:RD}, \(D\) has dimensions of length squared; the effective diffusivity \(D/\tau\) has dimensions of length squared per unit time.
	
	Following the tradition of mean-field reductions that capture the essential dynamics of complex spiking networks with minimal variables~\cite{wong2006}, we now coarse-grain the basal ganglia--thalamocortical circuit into two interacting layers. Let \(u_1(\mathbf{r},t)\) denote the activity of the upper layer (cortex--thalamus) and \(u_2(\mathbf{r},t)\) that of the lower layer (basal ganglia, dominated by STN and GPe). The interlayer interaction is modelled as a \emph{differential} coupling, i.e.\ the coupling force is proportional to the activity difference \((u_2-u_1)\) rather than to the absolute activity of either layer. Physiologically, this encodes the comparator function of the thalamus, which relays the mismatch between cortical and basal-ganglia signals rather than either signal alone. The resulting two-layer reaction-diffusion system is
	\begin{equation}\label{eq:two-layer}
			\begin{cases}
				\tau_1\,\frac{\partial u_1}{\partial t} &= -u_1 + f_1\!\left(w_{11}u_1 + D_1\nabla^2 u_1 + w_{12}(u_2-u_1) + I_1(t)\right), \\
				\tau_2\,\frac{\partial u_2}{\partial t} &= -u_2 + f_2\!\left(w_{22}u_2 + D_2\nabla^2 u_2 + w_{21}(u_1-u_2) + I_2(t)\right),
		\end{cases}
	\end{equation}
	with \(w_{11}\) and \(w_{22}\) the intra-layer effective couplings, \(w_{12}\) and \(w_{21}\) the interlayer differential couplings, \(D_{1,2}>0\) the diffusion coefficients inherited from the second moments of the corresponding kernels, and \(I_{1,2}(t)\) the external drives (e.g.\ DBS or sensory input). For the sigmoidal input-output functions we adopt
	\begin{equation}\label{eq:sigmoid}
		f_i(x) = \frac{f_{\max}^{(i)}}{1+\exp\!\left[-a_i\left(x-\theta_i\right)\right]},
		\qquad
		f_i'(x) = a_i f_i(x)\!\left[1-\frac{f_i(x)}{f_{\max}^{(i)}}\right] > 0,
	\end{equation}
	where \(f_{\max}^{(i)}>0\) is the saturation rate, \(a_i>0\) the gain, and \(\theta_i\) the half-activation threshold.
	
	Three features of Eq.~\eqref{eq:two-layer} are central to the analysis below. First, the diffusion terms are not postulated: they arise from the second moments of the nonlocal connectivity kernels, so that the spatial scale of any emergent pattern is fixed by the underlying neuroanatomy through \(D_1\) and \(D_2\). Second, the differential coupling enters linearly as \(w_{12}(u_2-u_1)\) and \(w_{21}(u_1-u_2)\), so that it vanishes identically when the two layers are synchronised. This provides a natural dynamical signature of pathological synchrony in Parkinson's disease, where the basal ganglia--thalamocortical loop is locked in the \(\beta\) band. Third, Eq.~\eqref{eq:two-layer} is the minimal extension of the single-layer reaction--diffusion model~\eqref{eq:RD} that retains both the nonlocal origin of the diffusion terms and the comparator structure of the interlayer interaction.				
	\subsection{Analysis of Turing instability}	
	The homogeneous steady state \((u_1^*,u_2^*)\) of Eq.~\eqref{eq:two-layer} satisfies
	\begin{equation}\label{eq:steady}
		\begin{aligned}
			u_1^* &= f_1\!\left(w_{11}u_1^* + w_{12}(u_2^*-u_1^*) + I_1\right), \\
			u_2^* &= f_2\!\left(w_{22}u_2^* + w_{21}(u_1^*-u_2^*) + I_2\right),
		\end{aligned}
	\end{equation}
	with \(I_{1,2}\) the time-averaged external drives. Eq.~\eqref{eq:steady} is solved numerically by Newton-Raphson iteration.

	Linearizing (see SM Sec.~\ref{sec:linearization}) Eq.~\eqref{eq:two-layer} about \((u_1^*,u_2^*)\) with perturbations \(\delta u_i(\mathbf{r},t)=c_i e^{\lambda t+i\mathbf{k}\cdot\mathbf{r}}\) yields
	\begin{equation}\label{eq:char}
		\tau_1\tau_2\lambda^2 + \bigl(\tau_1 A_2(k)+\tau_2 A_1(k)\bigr)\lambda
		+ A_1(k)A_2(k) - \beta_1\beta_2 w_{12}w_{21} = 0,
	\end{equation}
	where
	\begin{equation}\label{eq:Ak}
		A_1(k) = 1 - \beta_1(w_{11}-w_{12}) + \beta_1 D_1 k^2,
		\qquad
		A_2(k) = 1 - \beta_2(w_{22}-w_{21}) + \beta_2 D_2 k^2,
	\end{equation}
	and \(\beta_i=f_i'(x_i^*)>0\) is the steady-state gain, with \(x_i^*\) the total input at the fixed point. The differential coupling enters Eq.~\eqref{eq:Ak} only through the renormalized intra-layer couplings \(w_{11}-w_{12}\) and \(w_{22}-w_{21}\). This renormalization is the central structural effect of differential coupling: increasing \(w_{12}\) is equivalent to weakening the effective self-excitation of the upper layer, and similarly for \(w_{21}\) in the lower layer.
	
	In the absence of diffusion (\(D_1=D_2=0\)), \(A_i(0)\) are \(k\)-independent and the Routh--Hurwitz criterion (see SM Sec.~\ref{sec:turing}) requires
	\begin{equation}\label{eq:RH}
		T(0)\equiv\tau_1 A_2(0)+\tau_2 A_1(0)>0,
		\qquad
		D_{\det}(0)\equiv A_1(0)A_2(0)-\beta_1\beta_2 w_{12}w_{21}>0.
	\end{equation}
	When diffusion is switched on, a Turing instability occurs if there exists \(k>0\) such that \(\mathrm{Re}\,\lambda(k)>0\) while \(\mathrm{Re}\,\lambda(0)<0\). For the quadratic Eq.~\eqref{eq:char} this requires the determinant condition to fail while the trace remains positive,
	\begin{equation}\label{eq:turing}
		D_{\det}(k)\equiv A_1(k)A_2(k)-\beta_1\beta_2 w_{12}w_{21}<0,
		\qquad
		T(k)\equiv\tau_1 A_2(k)+\tau_2 A_1(k)>0.
	\end{equation}
	The first inequality is the Turing condition; the second excludes a Hopf-type instability. Writing \(F(y)\equiv A_1(k)A_2(k)-\beta_1\beta_2 w_{12}w_{21}\) with \(y=k^2\), Eq.~\eqref{eq:turing} becomes
	\begin{equation}\label{eq:F}
		F(y)=a y^2 + b y + c,
	\end{equation}
	with
	\begin{equation}\label{eq:abc}
		\begin{aligned}
			a &= \beta_1\beta_2 D_1 D_2 > 0, \\
			b &= \beta_1 D_1\bigl[1-\beta_2(w_{22}-w_{21})\bigr]
			+ \beta_2 D_2\bigl[1-\beta_1(w_{11}-w_{12})\bigr], \\
			c &= \bigl[1-\beta_1(w_{11}-w_{12})\bigr]\bigl[1-\beta_2(w_{22}-w_{21})\bigr]
			- \beta_1\beta_2 w_{12}w_{21}.
		\end{aligned}
	\end{equation}
	Since \(a>0\) and \(c>0\) by Eq.~\eqref{eq:RH}, \(F(y)\) is an upward parabola with \(F(0)=c>0\). A positive root with \(F(y)<0\) exists if and only if
	\begin{equation}\label{eq:turing-cond}
		\,b<0 \quad\text{and}\quad b^2>4ac\,.
	\end{equation}
	Condition~\eqref{eq:turing-cond} is the central result of the linear analysis: the first inequality ensures that the minimum of \(F\) lies at positive \(k^2\), the second that this minimum is negative. Physically, \(b<0\) requires the two renormalized intra-layer couplings to have opposite signs, and \(b^2>4ac\) requires the interlayer differential coupling to be strong enough to overcome the stabilizing intra-layer terms.
	
	At the Turing bifurcation, \(F(y)=F'(y)=0\), which yields the critical wavenumber
	\begin{equation}\label{eq:kc}
		k_c^2 = -\frac{b}{2a}
		= -\frac{
			\beta_1 D_1\bigl[1-\beta_2(w_{22}-w_{21})\bigr]
			+\beta_2 D_2\bigl[1-\beta_1(w_{11}-w_{12})\bigr]
		}{2\beta_1\beta_2 D_1 D_2}.
	\end{equation}
	The emergent pattern has spatial period \(\Lambda=2\pi/k_c\). Eq.~\eqref{eq:kc} makes explicit how the pattern scale is controlled jointly by the diffusion coefficients \(D_{1,2}\) and by the renormalized couplings \(w_{11}-w_{12}\) and \(w_{22}-w_{21}\): increasing \(D_2\) at fixed \(D_1\) lowers \(k_c\) and therefore widens the pattern, in agreement with the numerical observation that the stripe width grows monotonically with \(D_2\).	
	
	\subsection{Diverse pattern morphologies in the coupled neural field}	
\begin{figure}[ht]
	\centering
	\includegraphics[width=17cm]{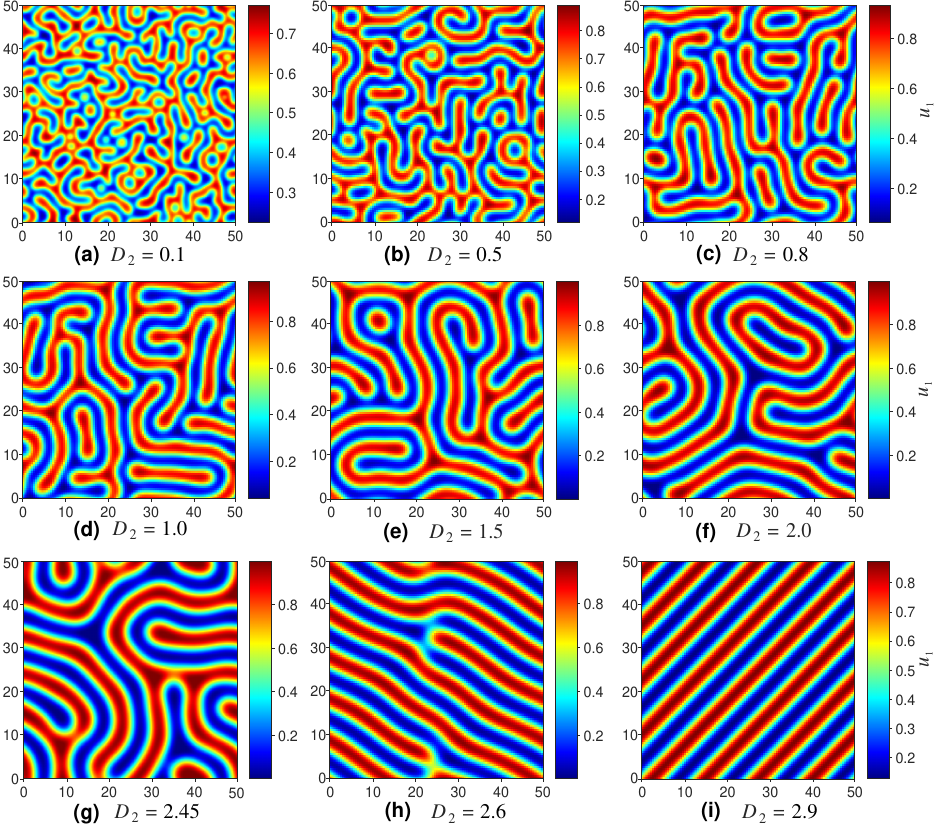}
	\caption{\textbf{Stripe patterns and the dependence of stripe width on \(D_2\).} Snapshots of \(u_1\) for increasing diffusion coefficient \(D_2 = 0.1, 0.5, 0.8, 1.0, 1.5, 2.0, 2.45, 2.6, 2.9\). As \(D_2\) increases, the stripe width \(W\) grows monotonically. Comparison between the theoretical prediction \(W = 2\pi\sqrt{D_2/(9-D_2)}\) derived from Eq.~\eqref{eq:kc} (solid line) and the numerical measurements obtained from the radial power spectrum (circles). The quantitative agreement validates the linear stability analysis. See SM Sec.~\ref{sec:Tablese} for simulation details.}
	\label{fig:stri_u1}
	\end{figure}

\begin{figure}[ht]
	\centering
	\includegraphics[width=17cm]{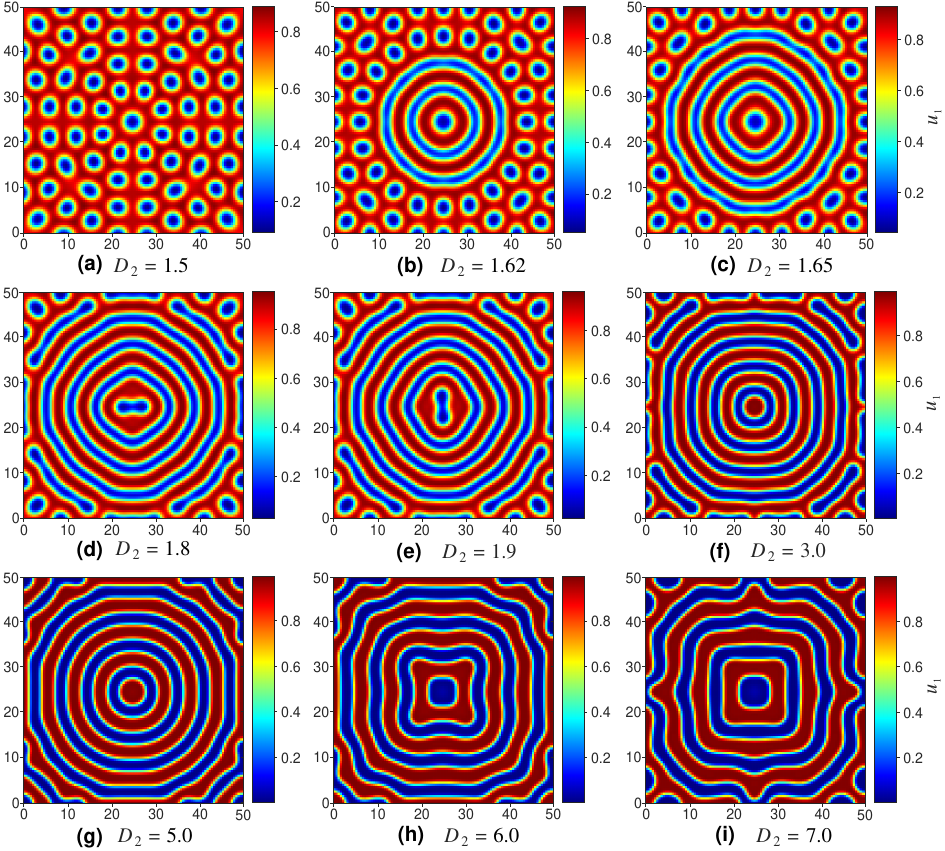}
	\caption{\textbf{Transition from spot arrays to concentric rings with increasing \(D_2\).} Snapshots of \(u_1\) for (a)--(c) \(D_2 = 1.5, 1.62, 1.65\) and (d)--(i) \(D_2 = 1.8, 1.9, 3.0, 5.0, 6.0, 7.0\). The transition from a hexagonal spot lattice to concentric rings is accompanied by a change in the spatial Fourier spectrum, reflecting the competition between the Turing wavelength \(\Lambda\) and the domain size. See SM Sec.~\ref{sec:Tablese} for simulation details.}
	\label{fig:dot_circle_u1}
	\end{figure}

\begin{figure}[ht]
		\centering
		\includegraphics[width=17cm]{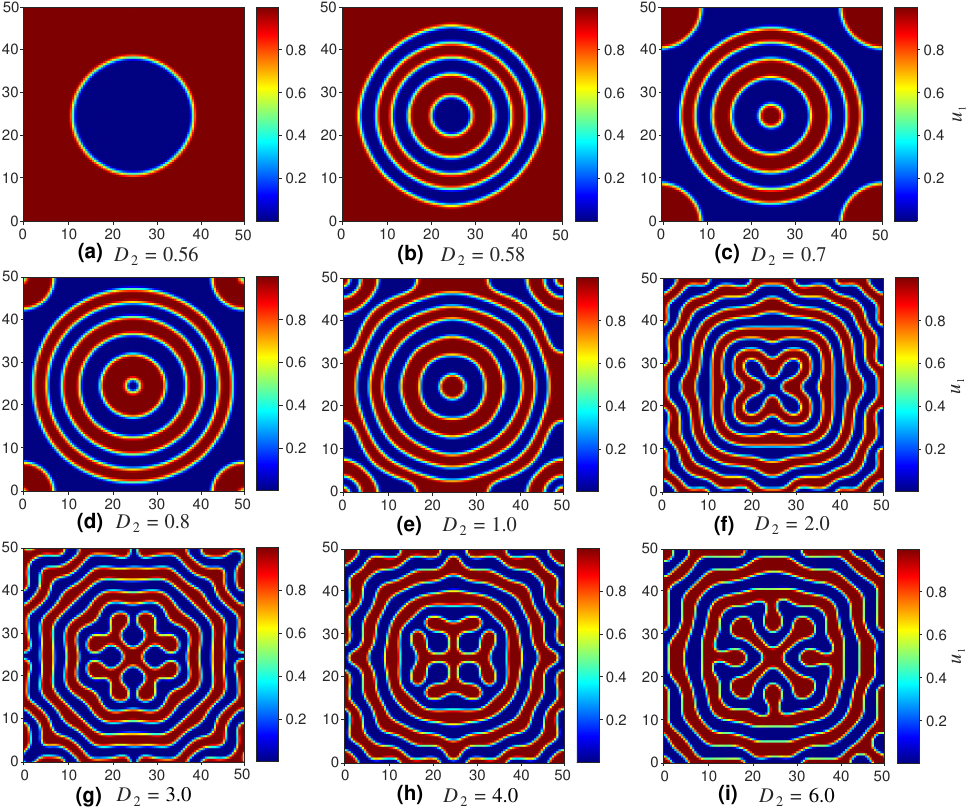}
		\caption{\textbf{Confinement-induced symmetry breaking of concentric rings.} Snapshots of \(u_1\) for increasing \(D_2 = 0.56, 0.58, 0.7, 0.8, 1.0, 2.0, 3.0, 4.0, 6.0\). As the Turing wavelength becomes comparable to the domain size, the circular ring patterns deform and acquire a fourfold symmetry (e.g., panels a--i), characteristic of finite-size Turing systems confined to a square domain. See SM Sec.~\ref{sec:Tablese} for simulation details.}
		\label{fig:babo_u1}
	\end{figure}

\begin{figure}[ht]
		\centering
		\includegraphics[width=17cm]{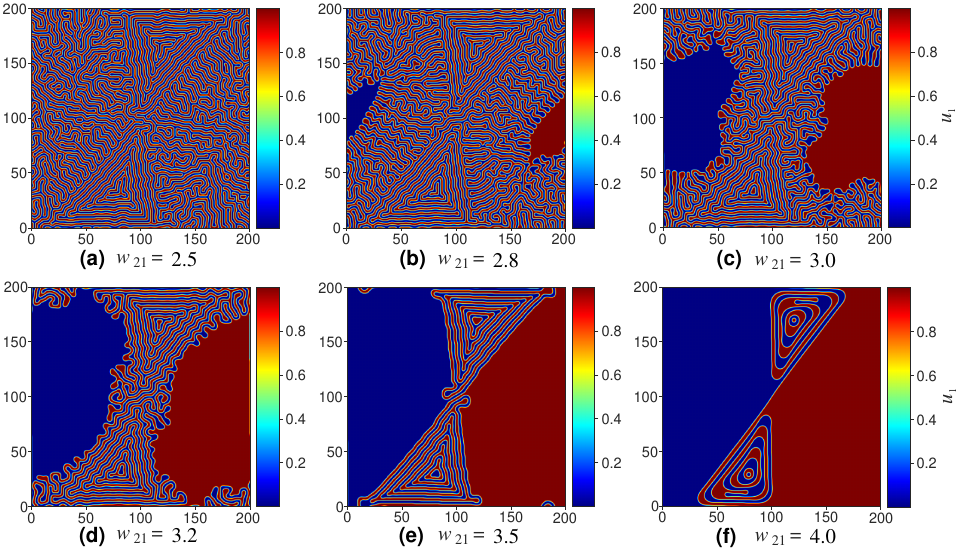}
		\caption{\textbf{Spiral waves and defect-mediated turbulence driven by differential coupling.} Snapshots of \(u_1\) for increasing interlayer coupling \(w_{21} = 2.5, 2.8, 3.0, 3.2, 3.5, 4.0\). For \(w_{21} \in [2.5, 3.2]\), the system displays defect-mediated turbulent states. Isolated spiral waves emerge for \(w_{21} \ge 3.5\). This regime is reached when the differential coupling drives the homogeneous state into a Hopf-unstable region, where the imaginary part of the eigenvalues dominates the dynamics. See SM Sec.~\ref{sec:Tablese} for simulation details.}
		\label{fig:Spiral_u1}
	\end{figure}

\begin{figure}[ht]
		\centering
		\includegraphics[width=17cm]{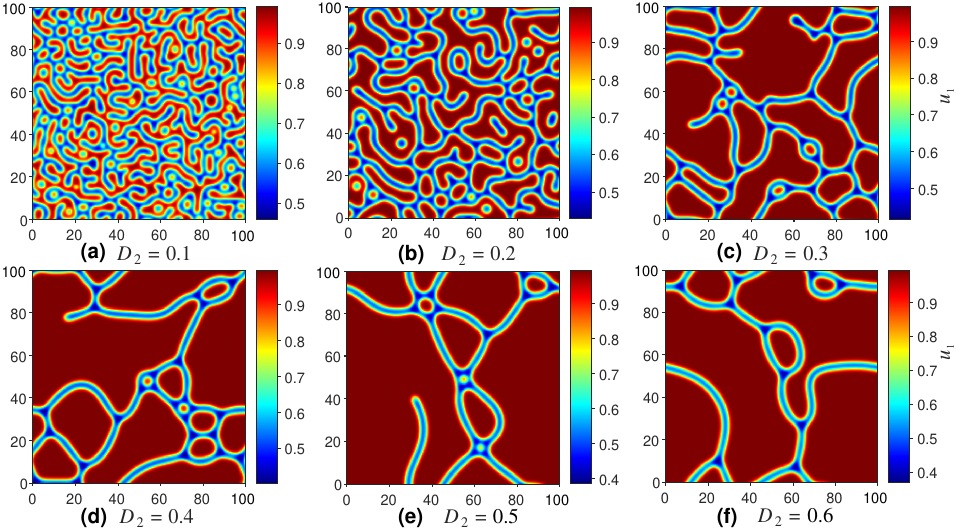}
		\caption{\textbf{Neuron-like branched patterns in the upper layer.} Spatial distribution of the activity \(u_1(\mathbf{r},t)\) in the steady regime, obtained from numerical integration of Eq.~\eqref{eq:two-layer} for a representative parameter set in the Turing-unstable region. The color bar indicates the dimensionless activity \(u_1\). The branched, dendrite-like structures emerge from the interaction between the sigmoid nonlinearity and the differential interlayer coupling. See SM Sec.~\ref{sec:Tablese} for simulation details.}
		\label{fig:neural_u1}
	\end{figure}

	The dispersion relation~\eqref{eq:char} fixes the range of unstable wavenumbers but not the selected morphology. The morphology is selected by the sigmoid nonlinearities and by the interlayer differential coupling. We integrate Eq.~\eqref{eq:two-layer} numerically on a periodic domain with a pseudospectral method, starting from a weak random perturbation of the homogeneous state. Noise-induced pattern formation in spatially extended neural media is a well-established mechanism by which stochastic fluctuations can select a characteristic spatial scale~\cite{perc2007coherence,perc2006diffusive}. In the present deterministic setting, the random perturbation serves only to seed the unstable modes selected by the dispersion relation; the asymptotic pattern is governed by the Turing condition rather than by the noise statistics. Five families of patterns are identified, shown in Fig.~\ref{fig:stri_u1}--\ref{fig:neural_u1}.
	\begin{figure}[ht]
	\centering
	\includegraphics[width=\columnwidth]{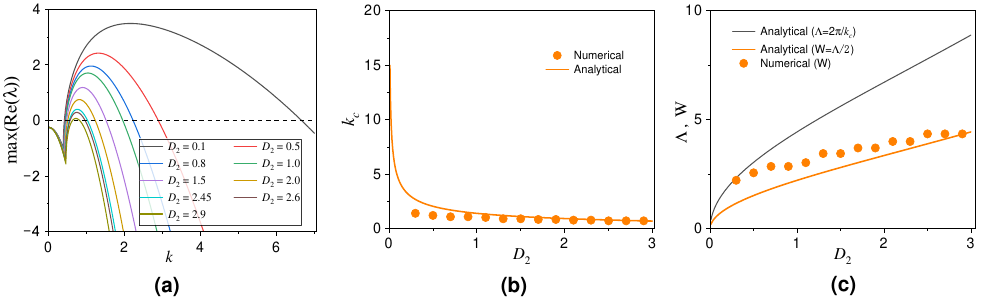}
	\caption{Dispersion relation, critical wavenumber, and stripe width of the stripe pattern.
		(a) Growth-rate dispersion curves for several values of \(D_2\). The unstable band narrows and the peak growth rate decreases as \(D_2\to3^-\). 
		(b) Critical wavenumber \(k_c\) versus \(D_2\): analytical prediction \(k_c=\sqrt{(9-D_2)/(4D_2)}\) (solid line) and values extracted from the simulated stripe patterns (dots). 
		(c) Stripe width \(W=\pi/k_c\) versus \(D_2\): analytical prediction \(W=2\pi\sqrt{D_2/(9-D_2)}\) (solid line) and measured widths (dots). 
		The numerical points in (b) and (c) fall on the analytical curves over the accessible range of \(D_2\), confirming that the stripe scale is selected by the linear dispersion relation. See SM Sec.~\ref{sec:Tablese} for simulation details.}
	\label{fig:sesan_tri}
	\end{figure}
	
	For $D_2<3$, the selected pattern is a family of parallel stripes (see Fig.~\ref{fig:stri_u1}). The dispersion curves in Fig.~\ref{fig:sesan_tri}(a) show that the band of unstable wavenumbers narrows as $D_2$ approaches $3$ from below, and that the maximum growth rate decreases monotonically over the same range. The stripe width is determined by the critical wavenumber in Eq.~\eqref{eq:kc}. Substituting the model parameters into this equation yields	
	\begin{equation}
		W(D_2)=\frac{\Lambda}{2}=\frac{\pi}{k_c(D_2)}
		=2\pi\sqrt{\frac{D_2}{9-D_2}},
		\label{eq:stripe-width}
	\end{equation}
	so that $W$ grows monotonically with $D_2$ and diverges as $D_2\to 3^-$. Fig.~\ref{fig:sesan_tri}(b) compares the analytical \(k_c(D_2)\) with the numerical solution obtained by solving the \(F(y)=F'(y)=0\), and Fig.~\ref{fig:sesan_tri}(c) compares the analytical $W(D_2)$ with the measured stripe width. In both panels the numerical points fall on the analytical curves over the accessible range of $D_2$, confirming that the stripe scale is selected by the linear dispersion relation. The stripe orientation, by contrast, is a nonlinear property that is not fixed by the linear analysis.
	
	The upper and lower layers share the same stripe morphology but differ in amplitude (see Fig.~\ref{fig:stri_u2}). The interlayer difference \(\Delta u=u_2-u_1\) decreases monotonically with increasing \(D_2\) (see Fig.~\ref{fig:stri_difference}). This is consistent with the dispersion relation~\eqref{eq:char}: as \(D_2\) grows, the Turing-unstable band \(k_{\rm on}<k<k_{\rm off}\) narrows and eventually collapses at \(D_2=3\), so that the two layers are driven toward a common homogeneous state. In the present two-layer model, a vanishing interlayer difference is the dynamical signature of pathological synchrony. Desynchronization of abnormally synchronized neuronal populations is the central therapeutic rationale of DBS in Parkinson's disease~\cite{tass2003}. The monotonic decrease of \(\Delta u\) with \(D_2\) therefore indicates that increasing the lower-layer diffusivity pushes the system toward a more synchronized, less therapeutically favorable regime.

	The spatial power spectrum (see SM Sec.~\ref{sec:Methods}) of \(u_1\) (see Fig.~\ref{fig:stri_power}) provides a quantitative measure of pattern order. For \(D_2<3\), the spectrum exhibits a broad ring, reflecting the excitation of a band of wavenumbers around the most unstable mode. As \(D_2\) approaches the Turing boundary, the ring narrows into a discrete set of peaks; at \(D_2=2.9\) only a few isolated peaks remain. This evolution is the spectral counterpart of the narrowing unstable band in Eq.~\eqref{eq:char}: the dominant wavenumber extracted from the power spectrum coincides with the linear prediction \(k_c(D_2)\) within the resolution of the simulation, and the wavelength \(2\pi/k_{\rm peak}\) reproduces the stripe width \(W\) in Eq.~\eqref{eq:stripe-width}. The emergence of sharp, isolated peaks is a general signature of well-selected Turing patterns with a characteristic wavelength independent of domain size~\cite{cross1993}, in contrast to the broad, continuous spectra of labyrinthine or defect-mediated states.
	
	In a separate parameter regime, the selected morphology is a hexagonal array of spots, which reorganises into concentric rings as \(D_2\) increases within that regime (see Fig.~\ref{fig:dot_circle_u1}). The dispersion curves in Fig.~\ref{fig:sesanSM}(a) show that the maximum real part of the growth rate remains positive over a broad band of wavenumbers for all values of \(D_2\) explored, from \(1.5\) to \(7.0\). The peak height increases monotonically with \(D_2\), so that the Turing instability persists throughout the spot-to-ring sequence. The spot-to-ring transition is accompanied by a change in the spatial power spectrum from a ring of peaks to a discrete set of peaks, reflecting the competition between the Turing wavelength and the domain size~\cite{cross1993} (see SM Sec.~\ref{sec:amplitude} for the nonlinear mode selection). The interlayer difference \(\Delta u\) remains confined to a narrow amplitude range throughout this transition (see Fig.~\ref{fig:dot_circle_difference}), indicating that the spot-to-ring rearrangement leaves the degree of interlayer synchrony essentially unchanged. This behaviour differs from the stripe regime, where \(\Delta u\) decreases monotonically with \(D_2\) (see Fig.~\ref{fig:stri_difference}).
		
	When the Turing wavelength becomes comparable to the domain size, the concentric rings acquire a fourfold-symmetric envelope (see Fig.~\ref{fig:babo_u1}). The dispersion curves in Fig.~\ref{fig:sesanSM}(b) show the peak height increasing as \(D_2\) grows. The fourfold-symmetric rings observed at large \(D_2\) therefore correspond to a nonlinear reorganization of an initially unstable state rather than to a persistent linear instability. This circular-to-square transition is typical of Turing systems confined to a square domain, where the boundary selects the lowest-order symmetric mode compatible with the domain geometry~\cite{cross1993}. In this regime the interlayer difference \(\Delta u\) increases by more than two orders of magnitude as \(D_2\) grows from \(0.56\) to \(6.0\) (see Fig.~\ref{fig:babo_difference}), so that increasing \(D_2\) progressively weakens interlayer synchrony. This trend is opposite to that of the stripe regime (see Fig.~\ref{fig:stri_difference}), where increasing \(D_2\) enhances synchrony. Because pathological \(\beta\)-band synchrony in Parkinson's disease is the target of desynchronizing deep brain stimulation~\cite{tass2003}, the opposite trends in the stripe and radial regimes imply that the therapeutic consequence of modulating the effective diffusion depends on which pattern morphology is selected.
	
	A different regime is reached when the differential coupling pushes the homogeneous state into a Hopf-unstable region. For \(2.5\le w_{21}\le4.0\), the system exhibits a sequence from a defect-mediated turbulent state, through a few spiral cores, to isolated spiral waves (see Fig.~\ref{fig:Spiral_u1}). The dispersion curves in Fig.~\ref{fig:sesanSM}(c) confirm that the instability changes character across this sequence. For $ 2.5\leq w_{21}\leq 4.0$ the growth rate at \(k=0\) is positive, so that the homogeneous state oscillates before spatial structure develops, consistent with the emergence of isolated spiral cores from a background of stable homogeneous activity. The transition is controlled by the ratio between the imaginary and real parts of the unstable eigenvalues, which measures the competition between oscillation and pattern formation~\cite{keener1998}. The coexistence of rotating and traveling waves, and noise-induced transitions between them, is a known feature of two-dimensional neural field models with competing excitation and inhibition~\cite{naoumenko2019waves}. The interlayer difference \(\Delta u\) decreases only by a factor of two across this sequence (see Fig.~\ref{fig:Spiral_difference}), so that the turbulence-to-spiral transition does not reflect a significant change in interlayer synchrony. For a spatially localized periodic drive at the center, the system develops concentric target waves. The emergence of target waves from a localized pacemaker is a generic feature of spatially extended excitable and oscillatory media~\cite{perc2009target}. The corresponding power spectrum (see Fig.~\ref{fig:Spiral_power}) changes from a broad ring to a small set of discrete peaks, consistent with the reduction of the number of active modes as the system moves from defect-mediated turbulence to isolated spiral cores.	
	
	At intermediate parameter values, the pattern develops branched, neuron-like morphologies with dendrite-like protrusions (see Fig.~\ref{fig:neural_u1}). The lower layer exhibits the same branched organisation (see Fig.~\ref{fig:neural_u2}), and both layers coarsen as \(D_2\) increases. The dispersion curves in Fig.~\ref{fig:sesanSM}(d) show that both the peak growth rate and the range of unstable wavenumbers decrease monotonically with increasing \(D_2\): the positive band is broadest at \(D_2=0.1\) and narrows progressively up to \(D_2=0.6\). Such structures are reminiscent of the morphologies obtained in reaction--diffusion models of biological branching~\cite{meinhardt1982,kondo2010}, but here arise from the sigmoid nonlinearity and the differential coupling rather than from any explicit branching mechanism. The interlayer difference \(\Delta u\) remains confined to a narrow amplitude range throughout this coarsening (see Fig.~\ref{fig:neural_difference}), so that the morphological reorganisation is not accompanied by a significant change in interlayer synchrony. This regime therefore occupies an intermediate position between the stripe regime, where \(\Delta u\) decreases monotonically with \(D_2\) (see Fig.~\ref{fig:stri_difference}), and the radial regime, where \(\Delta u\) grows by orders of magnitude (see Fig.~\ref{fig:babo_difference}).
	
	Taken together, the interlayer difference across the five regimes (see Figs.~\ref{fig:stri_difference}, \ref{fig:dot_circle_difference}, \ref{fig:babo_difference}, \ref{fig:Spiral_difference}, and~\ref{fig:neural_difference}) shows that the effect of a control parameter on interlayer synchrony is regime-dependent: it can be monotonic (stripes), negligible (spots and rings), strongly amplifying (radial modes), weak (spirals), or nearly conserved (neuron-like). This regime dependence is relevant to the interpretation of desynchronizing deep brain stimulation~\cite{tass2003}, because the same modulation may either enhance or weaken synchrony depending on the pattern morphology that the network selects.
	
	\subsection{Clinical mapping: beta synchronization and DBS}	
	The Turing patterns obtained above are stationary in time but structured in space, whereas clinical beta activity in Parkinson's disease is both temporally bursting and spatially organized. Recent intraoperative and chronic recordings provide quantitative anchors for comparison.
	
	Beta activity in the subthalamic nucleus (STN) of PD patients occurs in transient bursts rather than as a continuous oscillation. Long STN beta bursts ($>200$ ms) overlap temporally with cortical bursts in more than 30\% of cases, and cortical bursts precede STN onset by several milliseconds~\cite{tinkhauser2018}. The characteristic burst duration of 200--400 ms is an order of magnitude longer than the membrane time constants of individual neurons, and it sets a natural temporal scale for the integration process that culminates in a spatially structured state. In the present model, the time required for a Turing pattern to emerge from a weak perturbation is set by the inverse of the maximum growth rate $\lambda_{\max}(k_c)$; for the parameters of Fig.~\ref{fig:stri_u1}, $1/\lambda_{\max}\sim 10^2$--$10^3$ in dimensionless units, comparable to the observed burst duration when $\tau_1$ and $\tau_2$ are identified with synaptic time constants in the 10--100 ms range.
	
	The spatial extent of beta coherence in the STN and substantia nigra pars reticulata (SNr) has been measured directly: significant coherence between neuronal firing and LFPs persists at recording sites separated by approximately 1 mm, and levodopa administration greatly reduces this coherence~\cite{brown2001}. Beta bursts in the cortico--basal ganglia network form spatially discrete ensembles that can span non-contiguous pools of neurons within and across structures. The Turing wavelength $\Lambda=2\pi/k_c$ in Eq.~\eqref{eq:kc} defines the analogous spatial scale in the model: for the stripe regime, $\Lambda$ ranges from $\sim2$ to $\sim9$ in the same length units as $D_2^{1/2}$, and matching the observed $\sim$1 mm coherence length to the model wavelength yields an effective diffusion length $D_2^{1/2}\sim0.1$--$0.5$ mm. This is a semi-quantitative correspondence rather than a parameter fit, because the clinical coherence length is measured within a single nucleus whereas $\Lambda$ describes the full interlayer pattern. Nevertheless, both the clinical and model scales are set by a competition between local coupling and spatial spread, which is the defining feature of a Turing mechanism.
	
	Clinically effective STN-DBS acutely suppresses beta power in the STN and connected cortical regions, and the degree of suppression correlates with clinical improvement~\cite{kuhn2008}. DBS does not merely silence activity: it shifts STN firing from rhythmic synchrony to irregular, asynchronous discharge, consistent with a transition out of a synchronized attractor. In the model, DBS enters through $I_{1,2}(t)$ and displaces the operating point, thereby changing the gains $\beta_i$ and the coefficients $a,b,c$ in Eq.~\eqref{eq:abc}. When the displaced parameters no longer satisfy the Turing condition~\eqref{eq:turing-cond}, the pattern is destroyed and the system returns to the homogeneous state. The regime-dependent interlayer difference in Figs.~\ref{fig:stri_difference}, \ref{fig:dot_circle_difference}, \ref{fig:babo_difference}, \ref{fig:Spiral_difference}, and~\ref{fig:neural_difference} adds a qualification: a parameter change that suppresses one pattern morphology may not suppress another with the same efficacy, because the sign and magnitude of $\Delta u$ respond differently to the control parameter in different regimes. This is consistent with the clinical observation that the optimal DBS parameters are patient-specific and that beta suppression is not always a reliable predictor of motor improvement in every individual~\cite{kuhn2008}.
	
	\section{Discussions}		
	Beta-band synchronization in the basal ganglia and thalamocortical circuit is transient, spatially organized, and suppressed by effective subthalamic stimulation, yet the dynamical conditions that produce spatially structured synchrony and govern its response to parameter modulation remain poorly constrained by theory. The two-layer neural field model with differential coupling developed here relates these observations to a minimal set of dynamical ingredients. Its central structural feature is that the interlayer couplings $w_{12}$ and $w_{21}$ enter the dispersion relation in Eqs.~\eqref{eq:char} and \eqref{eq:kc} only through the renormalized combinations $w_{11}-w_{12}$ and $w_{22}-w_{21}$, and the Turing condition is met whenever the differential cross-terms $-\beta_1\beta_2 w_{12}w_{21}$ overcome the stabilizing intra-layer terms. This renormalization has a consequence that should be kept in view when pattern transitions are used to infer synaptic changes: at the linear level, a shift in interlayer coupling is indistinguishable from a shift in intra-layer self-excitation.
	
	The classical Wilson-Cowan model~\cite{wilson1972} and its nonlocal extensions~\cite{hutt2005} establish that reaction-diffusion dynamics can generate spatial structure, but they do not specify how two functionally distinct layers interact when the interaction is proportional to their activity difference. Phenomenological coupled-oscillator models of beta synchronization~\cite{tass2003} and reduced two-variable neural mass models derived from spiking networks~\cite{wong2006} capture complex temporal dynamics (such as synchrony transitions or perceptual decisions) but do not resolve the spatial scale. The present model retains the analytical tractability of these reductions while making the spatial scale of synchronization an explicit output.The differential coupling is not an additional phenomenological term; it follows from the comparator function of the thalamus. This highlights a universal principle in complex systems: local pairwise interactions, whether they manifest as pairwise encounters~\cite{kang2024self}, or higher-order interactions~\cite{kang2026higher}, can spontaneously drive macroscopic self-organization and pattern formation. 
		
	The morphological repertoire generated by the model (stripes, spots, rings, spiral waves, and branched structures) carries a dynamical message that is not evident from linear analysis alone. Since the seminal work of Turing~\cite{turing1952}, the formation of spatial structures from homogeneous states has been recognized as a generic dynamical phenomenon. However, the selection among different morphologies requires nonlinear analysis, which goes beyond the linear dispersion relation. The stripe width is quantitatively predicted by the model, and the remaining morphologies are selected by the sigmoid nonlinearity and by the competition between Turing and Hopf instabilities, analogous to morphology selection in biological pattern formation~\cite{cross1993,meinhardt1982,kondo2010}.  However, it is important to emphasize that this analogy is structural rather than mechanistic: the branched structures reported here arise from the sigmoid nonlinearity and the differential coupling, not from an explicit branching rule. The transition from defect-mediated turbulence to isolated spiral waves in the present model further resonates with the complex spatiotemporal dynamics observed in other nonlinear systems. For instance, homoclinic routes to chaos have been identified in omnivore communities~\cite{kang2026homoclinic}, and intraspecific interference has been shown to promote biodiversity through emergent nonlinear feedback~\cite{kang2024predator}. These parallels indicate that the mechanisms generating rich dynamical states are highly conserved across scales.
	
	The interlayer difference \(\Delta u\) provides a natural order parameter for interlayer synchrony, and its response to control parameters is strongly regime-dependent. In the stripe regime, \(\Delta u\) decreases monotonically with increasing diffusion, indicating that the two layers are driven toward synchrony. In the radial regime, the opposite occurs: \(\Delta u\) grows by orders of magnitude, driving the layers apart. In the spot-to-ring and spiral regimes, \(\Delta u\) remains confined or changes only weakly. This regime dependence is the central dynamical message of the present work. This behavior is consistent with our previous findings in coupled ecological systems, where self-organized pattern synchronization is strongly modulated by stochasticity and coupling strength~\cite{kang2026sync}. It implies that a single scalar index of synchrony, such as total beta power, cannot fully characterize the network response to a modulation: the same parameter shift can either enhance or weaken synchrony depending on which spatial morphology the network has selected.
	
	This network-level perspective is consistent with the emerging view that Parkinson's disease is a multi-cellular, spatially heterogeneous disease. Microglial iron efflux via LCN2 and oligodendrocyte IL-33 signaling are not isolated events; they alter the local metabolic and inflammatory environment~\cite{nus1_2026,oligo_2026}, which in turn can shift the excitability and coupling parameters of the neural field, thereby pushing the network into different pattern regimes. This provides a dynamical explanation for why a single therapeutic target (e.g., iron chelation or anti-inflammatory agents) might show variable efficacy across patients. This interpretation is consistent with the observation that chronic demyelination-associated microglial states are reversible by pharmacological intervention~\cite{demyel_2026}, and with the functional independence of basal ganglia neural populations encoding distinct variables~\cite{bg_speech_2026}.
	
	Within the present framework, DBS enters through the external drives \(I_{1,2}(t)\), which displace the operating point and therefore the gains \(\beta_i\) and the coefficients \(a,b,c\) in Eq.~\eqref{eq:abc}. When the displaced parameters no longer satisfy Eq.~\eqref{eq:turing-cond}, the pattern is destroyed and the homogeneous state is restored. Previous mean-field studies of the basal ganglia have shown that transmission delays critically shape the emergence of abnormal oscillations in the parkinsonian state~\cite{perc2024parkinson}, and that noise-induced synchronization can be suppressed by high-frequency periodic stimulation in a frequency range that matches clinical DBS settings~\cite{perc2020dbs}. The regime dependence of \(\Delta u\) adds a further constraint: the same displacement may suppress one morphology more efficiently than another, because the stability boundaries of different morphologies do not coincide. This provides a dynamical rationale for the patient-specific nature of optimal DBS parameters, and it suggests that stimulation efficacy should be assessed not only by beta suppression but also by the spatial morphology of the residual activity.
	
    \section{Conclusions}	
	We have derived a two-layer neural field model with differential interlayer coupling to describe the dynamics of the basal ganglia-thalamocortical circuit. The central finding is that differential coupling acts as a structural renormalization of the intra-layer dynamics: the interlayer couplings \(w_{12}\) and \(w_{21}\) enter the dispersion relation only through the combinations \(w_{11}-w_{12}\) and \(w_{22}-w_{21}\), providing a generic mechanism for Turing instability without requiring ad hoc cross-inhibition. The model generates a rich repertoire of spatially structured states, such as stripes, spots, rings, radial modes, spiral waves, and branched structures whose selection is determined by the sigmoid nonlinearity and the competition between Turing and Hopf instabilities. For the stripe regime, the analytical critical wavenumber accurately predicts the emergent wavelength, in quantitative agreement with the direct numerical solutions of the model equations.

	The interlayer difference \(\Delta u\) provides a natural order parameter for interlayer synchrony. We find that the response of \(\Delta u\) to parameter modulation is strongly regime-dependent: increasing the diffusion coefficient enhances synchrony in the stripe regime but weakens it in the radial regime, while leaving it nearly unchanged in the spot/ring and neuron-like regimes. This structural sensitivity challenges the sufficiency of single-scalar metrics (such as total beta power) for characterizing pathological states, and offers a dynamical rationale for the variable correlation between beta suppression and clinical improvement under deep brain stimulation (DBS). The results suggest that identifying the spatial morphology of the underlying network state, rather than relying solely on local oscillation amplitude, may offer a more informative target for optimizing stimulation parameters. Future work combining these theoretical predictions with clinical local field potential (LFP) recordings and interpretable machine learning is required to translate this framework into personalized DBS decision support.

	\section*{Acknowledgments}
	This study was funded by the Research Initiation Special Project of Guilin Medical University (20501026045).
	
	\section*{Conflict of interest}
	The authors declare that they have no known competing financial interests or personal relationships that could have appeared to influence the work reported in this paper.
	
	\section*{Author contributions}
	J.K. conceived the project and planned the study. All authors supervised the study, developed the model, analyzed the data, carried out the theoretical analysis and numerical simulations, analyzed the simulation results, and wrote the paper.
	
	\section*{Data and code availability}	
	%The data used in this study were obtained from previously published literature and publicly available databases, as cited in the main text. 
	The simulation code that supports the findings of this study is available from the corresponding author upon reasonable request.

	\newpage
	\textbf{\LARGE{Supplementary Materials}}
	
\renewcommand{\thesection}{\arabic{section}.}
\renewcommand{\thesubsection}{\arabic{section}.\arabic{subsection}}
\renewcommand{\thesubsubsection}{\arabic{section}.\arabic{subsection}.\arabic{subsubsection}}

\titleformat{\section}
{\fontsize{12}{9}\selectfont\bfseries\raggedright}
{\thesection}{0.3em}{}

\titleformat{\subsection}
{\fontsize{10}{9}\selectfont\bfseries\raggedright}
{\thesubsection}{0.3em}{}

\titleformat{\subsubsection}
{\fontsize{10}{9}\selectfont\bfseries\itshape\raggedright}
{\thesubsubsection}{0.3em}{}
	
	\makeatletter
	\renewcommand{\thefigure}{S\@arabic\c@figure}
	\renewcommand{\theequation}{S\@arabic\c@equation}
	\makeatother
	
	\setcounter{section}{0}
	\setcounter{figure}{0}
	\setcounter{equation}{0}
	\renewcommand\theequation{S\arabic{equation}}
	
	\section{Gradient expansion of the nonlocal convolution}
	\label{sec:expansion}
	%======================================================================
	
	The spatial convolution in Eq.~(2) of the main text reads
	\begin{equation}
		\mathcal{I}(\mathbf{r},t)=\int_{\Omega}w(\mathbf{r}-\mathbf{r}')\,u(\mathbf{r}',t)\,d\mathbf{r}'.
		\label{eq:S1}
	\end{equation}
	Setting $\boldsymbol{\xi}=\mathbf{r}-\mathbf{r}'$ and extending the integration domain to $\mathbb{R}^d$ (valid when the kernel range is much smaller than the system size),
	\begin{equation}
		\mathcal{I}(\mathbf{r},t)=\int_{\mathbb{R}^d}w(\boldsymbol{\xi})\,u(\mathbf{r}-\boldsymbol{\xi},t)\,d\boldsymbol{\xi}.
		\label{eq:S2}
	\end{equation}
	Expanding $u(\mathbf{r}-\boldsymbol{\xi},t)$ about $\mathbf{r}$,
	\begin{equation}
		u(\mathbf{r}-\boldsymbol{\xi},t)
		=u-\sum_{i}\xi_i\partial_i u+\frac{1}{2}\sum_{i,j}\xi_i\xi_j\partial_i\partial_j u
		-\frac{1}{6}\sum_{i,j,k}\xi_i\xi_j\xi_k\partial_i\partial_j\partial_k u+\mathcal{O}(|\boldsymbol{\xi}|^4),
		\label{eq:S3}
	\end{equation}
	with $u\equiv u(\mathbf{r},t)$ and $\partial_i\equiv\partial/\partial r_i$.
	
	\textit{Zeroth order.} With $W_0\equiv\int w(\boldsymbol{\xi})\,d\boldsymbol{\xi}$,
	\begin{equation}
		\int w(\boldsymbol{\xi})\,u\,d\boldsymbol{\xi}=W_0\,u.
		\label{eq:S4}
	\end{equation}
	
	\textit{First order.} Since $w(\boldsymbol{\xi})=w(-\boldsymbol{\xi})$ is even, $\xi_i w(\boldsymbol{\xi})$ is odd in $\xi_i$, hence
	\begin{equation}
		\int w(\boldsymbol{\xi})\,\xi_i\,d\boldsymbol{\xi}=0.
		\label{eq:S5}
	\end{equation}
	The first-order contribution vanishes.
	
	\textit{Second order.} For an isotropic kernel $w(\boldsymbol{\xi})=w(|\boldsymbol{\xi}|)$, the second moment tensor is isotropic:
	\begin{equation}
		\int w(\boldsymbol{\xi})\,\xi_i\xi_j\,d\boldsymbol{\xi}=\delta_{ij}\int w(\boldsymbol{\xi})\,\xi_1^2\,d\boldsymbol{\xi}.
		\label{eq:S6}
	\end{equation}
	Defining the scalar second moment
	\begin{equation}
		W_2\equiv\int|\boldsymbol{\xi}|^2w(\boldsymbol{\xi})\,d\boldsymbol{\xi}=d\int\xi_1^2w(\boldsymbol{\xi})\,d\boldsymbol{\xi},
		\label{eq:S7}
	\end{equation}
	the second-order term becomes
	\begin{equation}
		\frac{1}{2}\int w(\boldsymbol{\xi})\,\xi_1^2\,d\boldsymbol{\xi}\sum_i\partial_i^2u
		=\frac{W_2}{2d}\nabla^2u.
		\label{eq:S8}
	\end{equation}
	
	\textit{Third order.} The product $\xi_i\xi_j\xi_k w(\boldsymbol{\xi})$ is odd under $\boldsymbol{\xi}\to-\boldsymbol{\xi}$ (at least one index appears an odd number of times), so the integral vanishes.
	
	\textit{Higher orders.} The remainder is $\mathcal{O}(|\boldsymbol{\xi}|^4)$, negligible when the kernel range is small compared with the spatial scale of $u$.
	
	Collecting terms,
	\begin{equation}
		\mathcal{I}(\mathbf{r},t)=W_0\,u+D\,\nabla^2u+\mathcal{O}(|\boldsymbol{\xi}|^4),\qquad D\equiv\frac{W_2}{2d}.
		\label{eq:S9}
	\end{equation}
	
	\textit{Dimensions.} In Eq.~(16) of the main text, $\nabla^2$ carries dimensions of $[\text{length}]^{-2}$ and $u$ is dimensionless, so $[D]=[\text{length}]^2$. The effective diffusivity $\tilde D=D/\tau$ has dimensions $[\text{length}]^2/[\text{time}]$. The condition $D>0$ is equivalent to $W_2>0$, i.e., the kernel has a net positive spatial spread.
	
	%======================================================================
	\section{Linearization and characteristic equation}
	\label{sec:linearization}
	%======================================================================
	
	\subsection{Steady state}
	
	The homogeneous steady state $(u_1^*,u_2^*)$ satisfies
	\begin{equation}
		\begin{cases}
			u_1^*=f_1\bigl(w_{11}u_1^*+w_{12}(u_2^*-u_1^*)+I_1\bigr),\\
			u_2^*=f_2\bigl(w_{22}u_2^*+w_{21}(u_1^*-u_2^*)+I_2\bigr).
		\end{cases}
		\label{eq:S10}
	\end{equation}
	The total inputs at the fixed point are
	\begin{equation}
		x_1^*\equiv w_{11}u_1^*+w_{12}(u_2^*-u_1^*)+I_1,\qquad
		x_2^*\equiv w_{22}u_2^*+w_{21}(u_1^*-u_2^*)+I_2.
		\label{eq:S11}
	\end{equation}
	
	\subsection{Perturbation expansion}
	
	Writing $u_i=u_i^*+\delta u_i$ with $|\delta u_i|\ll1$ and expanding the sigmoid,
	\begin{equation}
		f_i(x_i^*+\delta x_i)=f_i(x_i^*)+\beta_i\,\delta x_i+\tfrac{1}{2}\gamma_i(\delta x_i)^2+\mathcal{O}(|\delta x_i|^3),
		\label{eq:S12}
	\end{equation}
	where
	\begin{equation}
		\beta_i\equiv f_i'(x_i^*),\qquad \gamma_i\equiv f_i''(x_i^*).
		\label{eq:S13}
	\end{equation}
	For the logistic sigmoid, Eq.~(18) of the main text,
	\begin{align}
		f_i'(x)&=\frac{a_i f_{\max}^{(i)}e^{-a_i(x-\theta_i)}}{[1+e^{-a_i(x-\theta_i)}]^2},\label{eq:S14}\\
		f_i''(x)&=\frac{a_i^2f_{\max}^{(i)}e^{-a_i(x-\theta_i)}\bigl[1-e^{-a_i(x-\theta_i)}\bigr]}{[1+e^{-a_i(x-\theta_i)}]^3}.\label{eq:S15}
	\end{align}
	At $x=\theta_i$, $f_i''(\theta_i)=0$ and $\beta_i=a_i f_{\max}^{(i)}/4$.
	
	\subsection{Linearized system}
	
	Linearizing the coupling terms,
	\begin{equation}
		w_{12}(u_2-u_1)=w_{12}(u_2^*-u_1^*)+w_{12}(\delta u_2-\delta u_1),
		\label{eq:S16}
	\end{equation}
	and similarly for $w_{21}(u_1-u_2)$, one obtains
	\begin{equation}
		\begin{cases}
			\tau_1\partial_t\delta u_1=\bigl[-1+\beta_1(w_{11}-w_{12})+\beta_1D_1\nabla^2\bigr]\delta u_1+\beta_1w_{12}\delta u_2,\\[2pt]
			\tau_2\partial_t\delta u_2=\beta_2w_{21}\delta u_1+\bigl[-1+\beta_2(w_{22}-w_{21})+\beta_2D_2\nabla^2\bigr]\delta u_2.
		\end{cases}
		\label{eq:S17}
	\end{equation}
	
	\subsection{Fourier modes and characteristic equation}
	
	Substituting $\delta u_i=c_i e^{\lambda t+i\mathbf{k}\cdot\mathbf{r}}$ into Eq.~\eqref{eq:S17},
	\begin{equation}
		\begin{cases}
			\bigl[\tau_1\lambda+A_1(k)\bigr]c_1-\beta_1w_{12}c_2=0,\\
			-\beta_2w_{21}c_1+\bigl[\tau_2\lambda+A_2(k)\bigr]c_2=0,
		\end{cases}
		\label{eq:S18}
	\end{equation}
	where
	\begin{equation}
		A_1(k)\equiv1-\beta_1(w_{11}-w_{12})+\beta_1D_1k^2,\qquad
		A_2(k)\equiv1-\beta_2(w_{22}-w_{21})+\beta_2D_2k^2.
		\label{eq:S19}
	\end{equation}
	Nontrivial solutions require
	\begin{equation}
		\tau_1\tau_2\lambda^2+\bigl(\tau_1A_2+\tau_2A_1\bigr)\lambda+A_1A_2-\beta_1\beta_2w_{12}w_{21}=0,
		\label{eq:S20}
	\end{equation}
	which is Eq.~(30) of the main text.
	
	%======================================================================
	\section{Routh--Hurwitz conditions and the Turing criterion}
	\label{sec:turing}
	%======================================================================
	
	\subsection{Diffusionless stability}
	
	Setting $D_1=D_2=0$ gives $A_i(0)=1-\beta_i(w_{ii}-w_{ij})$, and Eq.~\eqref{eq:S20} reduces to
	\begin{equation}
		\tau_1\tau_2\lambda^2+T(0)\lambda+D_{\det}(0)=0,
		\label{eq:S21}
	\end{equation}
	with
	\begin{equation}
		T(0)\equiv\tau_1A_2(0)+\tau_2A_1(0),\qquad
		D_{\det}(0)\equiv A_1(0)A_2(0)-\beta_1\beta_2w_{12}w_{21}.
		\label{eq:S22}
	\end{equation}
	For a quadratic $a\lambda^2+b\lambda+c$ with $a>0$, both roots have negative real parts if and only if $b>0$ and $c>0$. Hence the diffusionless state is stable when
	\begin{equation}
		T(0)>0,\qquad D_{\det}(0)>0.
		\label{eq:S23}
	\end{equation}
	
	\subsection{Diffusion-induced instability}
	
	When $D_1,D_2>0$, $A_i(k)$ increase with $k$. A Turing instability occurs if there exists $k>0$ such that
	\begin{equation}
		D_{\det}(k)\equiv A_1(k)A_2(k)-\beta_1\beta_2w_{12}w_{21}<0,
		\label{eq:S24}
	\end{equation}
	while $T(k)>0$. The constant term of Eq.~\eqref{eq:S20} is then negative, so one eigenvalue is real and positive.
	
	\subsection{Parabolic representation}
	
	Let $y=k^2$. Then $D_{\det}(y)=F(y)\equiv ay^2+by+c$, with
	\begin{equation}
		a=\beta_1\beta_2D_1D_2>0,\qquad
		b=\beta_1D_1A_2(0)+\beta_2D_2A_1(0),\qquad
		c=A_1(0)A_2(0)-\beta_1\beta_2w_{12}w_{21}.
		\label{eq:S25}
	\end{equation}
	By Eq.~\eqref{eq:S23}, $c>0$. Since $a>0$, $F(y)$ is an upward parabola with $F(0)=c>0$. A positive root $y>0$ with $F(y)<0$ exists if and only if
	\begin{equation}
		\Delta=b^2-4ac>0,\qquad y_0=-\frac{b}{2a}>0,
		\label{eq:S26}
	\end{equation}
	i.e.,
	\begin{equation}
		b<0\qquad\text{and}\qquad b^2>4ac.
		\label{eq:S27}
	\end{equation}
	This is the Turing condition, Eq.~(42) of the main text.
	
	\subsection{Critical wavenumber}
	
	At the bifurcation, $\Delta=0$ and $F(y)$ is tangent to the $y$-axis at
	\begin{equation}
		y_c=-\frac{b}{2a}.
		\label{eq:S28}
	\end{equation}
	Hence
	\begin{equation}
		k_c^2=-\frac{b}{2a}
		=-\frac{\beta_1D_1A_2(0)+\beta_2D_2A_1(0)}{2\beta_1\beta_2D_1D_2},
		\label{eq:S29}
	\end{equation}
	which is Eq.~(43) of the main text.
	
	\subsection{Trace condition at finite $k$}
	
	From Eq.~\eqref{eq:S19},
	\begin{equation}
		T(k)=\tau_1A_2(k)+\tau_2A_1(k)=T(0)+\bigl(\tau_1\beta_2D_2+\tau_2\beta_1D_1\bigr)k^2.
		\label{eq:S30}
	\end{equation}
	Since all parameters are positive, $T(k)$ increases monotonically with $k^2$. If $T(0)>0$, then $T(k)>0$ for all $k>0$. Consequently, in the present model, a diffusion-driven instability (from $D_{\det}<0$) cannot coexist with a Hopf instability (from $T(k)<0$); whenever the diffusionless state is stable, any spatial instability is of pure Turing type.
	
	%======================================================================
	\section{Amplitude equation and nonlinear selection}
	\label{sec:amplitude}
	%======================================================================
	
	\subsection{Quadratic nonlinearity}
	
	Near the critical wavenumber $k_c$, write the perturbation as
	\begin{equation}
		\delta u_i=\epsilon\bigl[A(T)e^{ik_cx}+\bar A(T)e^{-ik_cx}\bigr]\phi_i+\mathcal{O}(\epsilon^2),
		\label{eq:S31}
	\end{equation}
	with slow time $T=\epsilon^2t$, linear eigenvector components $\phi_i$, and complex amplitude $A$. Substituting into Eq.~\eqref{eq:S17} and retaining terms up to $\mathcal{O}(\epsilon^3)$ yields the amplitude equation
	\begin{equation}
		\partial_TA=\mu A+\nu\bar A^2-g|A|^2A+\mathcal{O}(|A|^4),
		\label{eq:S32}
	\end{equation}
	where $\mu\propto(k_c^2-k^2)$ measures the distance from the bifurcation, $\nu$ is the quadratic coefficient, and $g$ is the cubic coefficient.
	
	The quadratic coefficient $\nu$ originates from the second derivative $\gamma_i=f_i''(x_i^*)$. When $x_i^*=\theta_i$, $\gamma_i=0$ and $\nu=0$: the bifurcation is supercritical and the selected pattern is stripes. When $x_i^*\neq\theta_i$, $\gamma_i\neq0$ and $\nu\neq0$: the bifurcation is subcritical, and hexagonal patterns may appear.
	
	\subsection{Explicit form of the quadratic term}
	
	At second order, the perturbation satisfies
	\begin{equation}
		\mathcal{L}\begin{pmatrix}\delta u_1^{(2)}\\\delta u_2^{(2)}\end{pmatrix}
		=-\frac{1}{2}\begin{pmatrix}\beta_1\gamma_1(\delta x_1)^2\\[2pt]\beta_2\gamma_2(\delta x_2)^2\end{pmatrix},
		\label{eq:S33}
	\end{equation}
	where $\mathcal{L}$ is the linear operator from Eq.~\eqref{eq:S17}. Since $\delta x_i$ contains $e^{\pm ik_cx}$, the square contains $e^{\pm2ik_cx}$ and constant terms. The second-order solution therefore contains $2k_c$ modes and zero modes. At third order, the interaction of the $2k_c$ mode with the $k_c$ mode produces a $k_c$ contribution whose coefficient is $\nu$.
	
	For the present model, the sign and magnitude of $\nu$ are determined by $\gamma_i$, $\beta_i$, $w_{ij}$, and $D_i$. When $\gamma_1$ and $\gamma_2$ have the same sign and comparable magnitude, $\nu$ is significantly nonzero; when $\gamma_1\approx-\gamma_2$, $\nu$ may cancel.
	
	\subsection{Stripe--hexagon criterion}
	
	For $\nu=0$ (supercritical), Eq.~\eqref{eq:S32} reduces to
	\begin{equation}
		\partial_TA=\mu A-g|A|^2A,
		\label{eq:S34}
	\end{equation}
	with the stable solution $|A|=\sqrt{\mu/g}$, corresponding to stripes.
	
	For $\nu\neq0$ (subcritical), stable hexagonal solutions exist when $\nu^2/(4g)>\mu>0$, i.e., when the system is in the bistable region. This is consistent with the statement in the main text that $\gamma_i\neq0$ is a necessary condition for hexagons.
	
	%======================================================================
	\section{Methods}
	\label{sec:Methods}
	%\subsection{Pseudospectral scheme}	
	The two-dimensional PDE (17) of the main text is solved on a periodic domain $[0,L_x]\times[0,L_y]$ with $N_x\times N_y$ grid points, $x_j=j\Delta x$, $y_l=l\Delta y$, $\Delta x=L_x/N_x$, $\Delta y=L_y/N_y$. The Laplacian is evaluated via the two-dimensional FFT,
	\begin{equation}
		\nabla^2u(x_j,y_l)=\mathcal{F}^{-1}\bigl[-k^2\,\mathcal{F}[u]\bigr]_{j,l},
		\label{eq:S35}
	\end{equation}
	where $\mathcal{F}$ denotes the discrete Fourier transform and
	\begin{equation}
		k_x=\frac{2\pi}{L_x}\bigl[0,1,\dots,N_x/2-1,-N_x/2,\dots,-1\bigr],
		\label{eq:S36}
	\end{equation}
	with $k_y$ defined analogously.
	
	%\subsection{Time integration}	
	Explicit Euler integration,
	\begin{equation}
		u_i^{n+1}=u_i^n+\Delta t\,\frac{-u_i^n+f_i(x_i^n)}{\tau_i},
		\label{eq:S37}
	\end{equation}
	is used with time step satisfying
	\begin{equation}
		\Delta t<\min\Bigl(\frac{\tau_1}{2},\frac{\tau_2}{2},\frac{\Delta x^2}{4D_1},\frac{\Delta y^2}{4D_2}\Bigr).
		\label{eq:S38}
	\end{equation}
	The production runs use $\Delta t=0.001$, $\Delta x=\Delta y\approx0.39$, $D_1=5$, and $D_2\le3$, satisfying Eq.~\eqref{eq:S38}.	
	
%	\subsection{Power spectrum}	
	The spatial power spectrum is
	\begin{equation}
		P(\mathbf{k})=\bigl|\mathcal{F}[u_1-\langle u_1\rangle]\bigr|^2,
		\label{eq:S39}
	\end{equation}
	with $\langle u_1\rangle$ the spatial average. The radial spectrum is obtained by binning in $|\mathbf{k}|$:
	\begin{equation}
		P_{\rm rad}(k_m)=\frac{1}{N_m}\sum_{|\mathbf{k}|\in[k_m,k_{m+1})}P(\mathbf{k}),
		\label{eq:S40}
	\end{equation}
	where $N_m$ is the number of modes in the $m$-th radial bin. The peak wavenumber $k_{\rm peak}$ is the location of the maximum of $P_{\rm rad}$, excluding the zero-frequency mode.
		
	\section{Supplemental Tables}\label{sec:Tablese}
	
	\begin{table}[htbp]
	\centering
	\caption{Parameter values used in the simulations of Figs.~\ref{fig:sesan_tri}-\ref{fig:neural_u1}.}
	\label{tab:params_row}
	\resizebox{\textwidth}{!}{%
		\begin{tabular}{@{}cccccccccccccc@{}}
			\toprule
			 {Figure} &   \multicolumn{11}{c}{Parameter values} \\
			\cmidrule(lr){2-14}
			& \(\tau_1\) & \(\tau_2\) & \(D_1\) & \(D_2\) & \(w_{11}\) & \(w_{22}\) & \(w_{12}\) & \(w_{21}\) & \(I_1\) & \(I_2\) & \(a_{1,2}\) & \(\theta_{1,2}\) & \(f_{\max}^{(1,2)}\) \\
			\midrule
			Figs.~\ref{fig:sesan_tri},~\ref{fig:stri_u1}   & 1.0 & 2.0 & 5.0 & [0.1,2.9]  & 3.0  & 2.0  & 5.0  & \(-3.0\) & \(-1.1\)  & \(-0.6\)  & 8.0 & 0.4 & 1.0 \\
			
			Fig.~\ref{fig:dot_circle_u1} & 1.0 & 0.5 & 0.3 & [1.5,7.0] & \(-0.5\) & \(-0.5\) & \(-2.5\) & 2.0  & 0.845 & 0.845 & 8.0 & 0.4 & 1.0 \\
			
			Fig.~\ref{fig:babo_u1} & 2.9 & 1.0 & 1.0 & [0.56,6.0] & 3 & 2 & \(-5\) & 4.0  & $-1.1$ & $-0.6$ & 8.0 & 0.4 & 1.0 \\

			Fig.~\ref{fig:Spiral_u1} & 2.9 & 1.0 & 1.0 & 1.62 & 3 & 2 & \(-5\) & [2.5,4.0]  & $-1.1$ & $-0.6$ & 8.0 & 0.4 & 1.0 \\			
			
			Fig.~\ref{fig:neural_u1} & 1.0 & 3 & 2 & [0.1,0.6] & \(-1\) & 3 & \(-1\) & 1.0  & 1.0506 & $-1.3494$ & 8.0 & 0.4 & 1.0 \\			
			\bottomrule								
		\end{tabular}%
	}
	\end{table}

	\clearpage
	\section*{Supplementary Figures}		
\begin{figure}[ht!]
	\centering
	\includegraphics[width=\columnwidth]{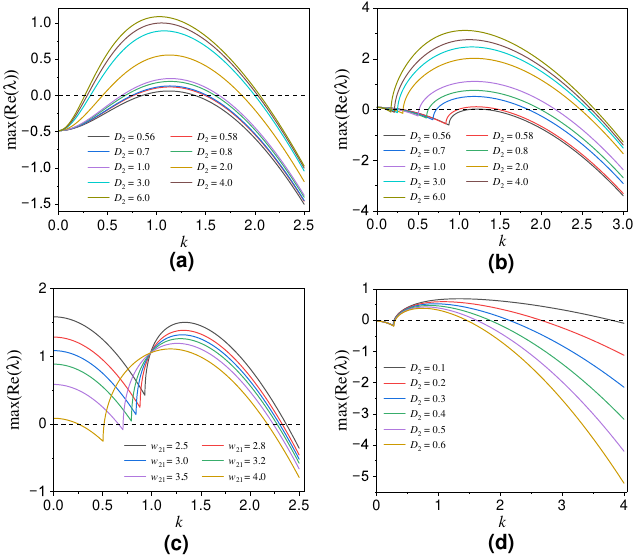}
	\caption{Growth-rate dispersion curves for the four pattern regimes. 
		(a) Spots and rings (Fig.~\ref{fig:dot_circle_u1}): the maximum real part is positive over a broad band for all \(D_2=1.5\)--\(7.0\), and the peak height decreases monotonically with \(D_2\). 
		(b) Radial modes (Fig.~\ref{fig:babo_u1}): a positive finite-\(k\) maximum is present for \(D_2\le1.0\); for \(D_2\ge2.0\) the maximum real part falls to zero or below, so that the large-\(D_2\) rings are nonlinearly sustained. 
		(c) Spiral waves (Fig.~\ref{fig:Spiral_u1}): for \(w_{21}\le3.0\) the growth rate at \(k=0\) is positive (Hopf instability); for \(w_{21}\ge3.2\) the \(k=0\) mode is stable and a narrow finite-\(k\) band remains. 
		(d) Neuron-like branched structures (Fig.~\ref{fig:neural_u1}): a small positive maximum is present only for \(D_2=0.1\) and \(0.2\); for \(D_2\ge0.3\) the maximum real part is negative. 
		The horizontal dashed line marks zero growth rate. Parameters not specified in the panels are as in the corresponding main-text figures.}
	\label{fig:sesanSM}
\end{figure}
		
	\clearpage
\begin{figure}[ht!]
	\centering
	\includegraphics[width=\columnwidth]{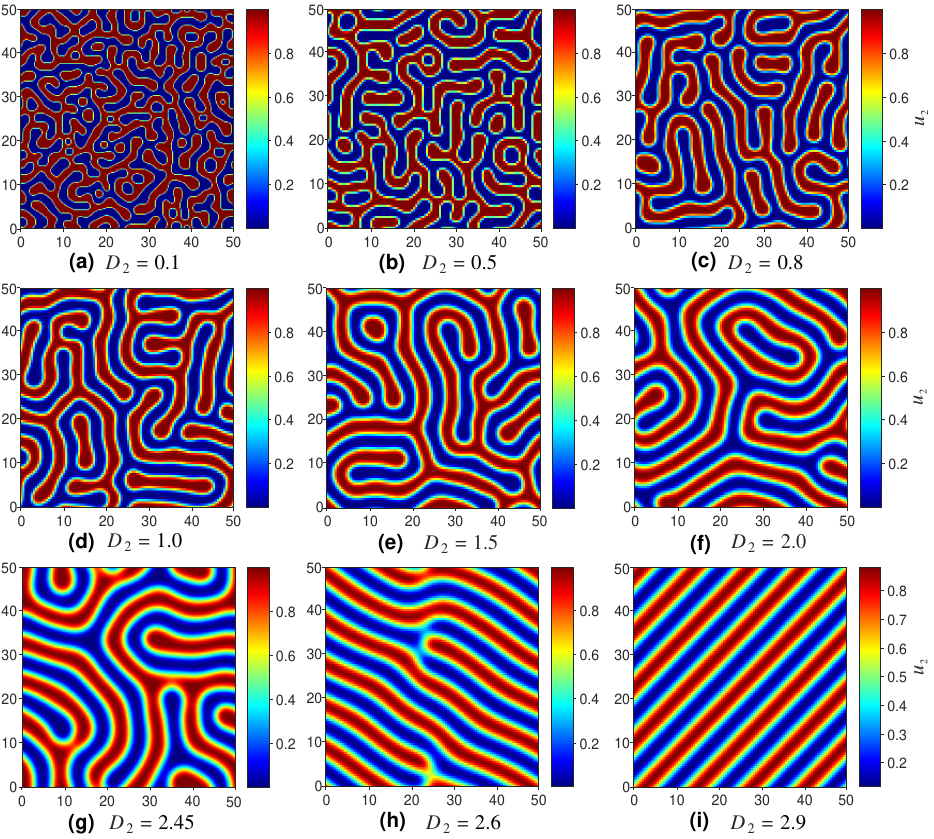}
	\caption{Stripe morphology of the lower layer \(u_2(\mathbf{r},t)\) for increasing \(D_2\). Snapshots for \(D_2=0.1,0.5,0.8,1.0,1.5,2.0,2.45,2.6,2.9\). The spatial structure of \(u_2\) mirrors that of \(u_1\), but the amplitude contrast between the two layers decreases with increasing \(D_2\), anticipating the reduction of the interlayer difference shown in Fig.~\ref{fig:stri_difference}.}
	\label{fig:stri_u2}
\end{figure}
	\clearpage
\begin{figure}[htbp]
	\centering
	\includegraphics[width=\columnwidth]{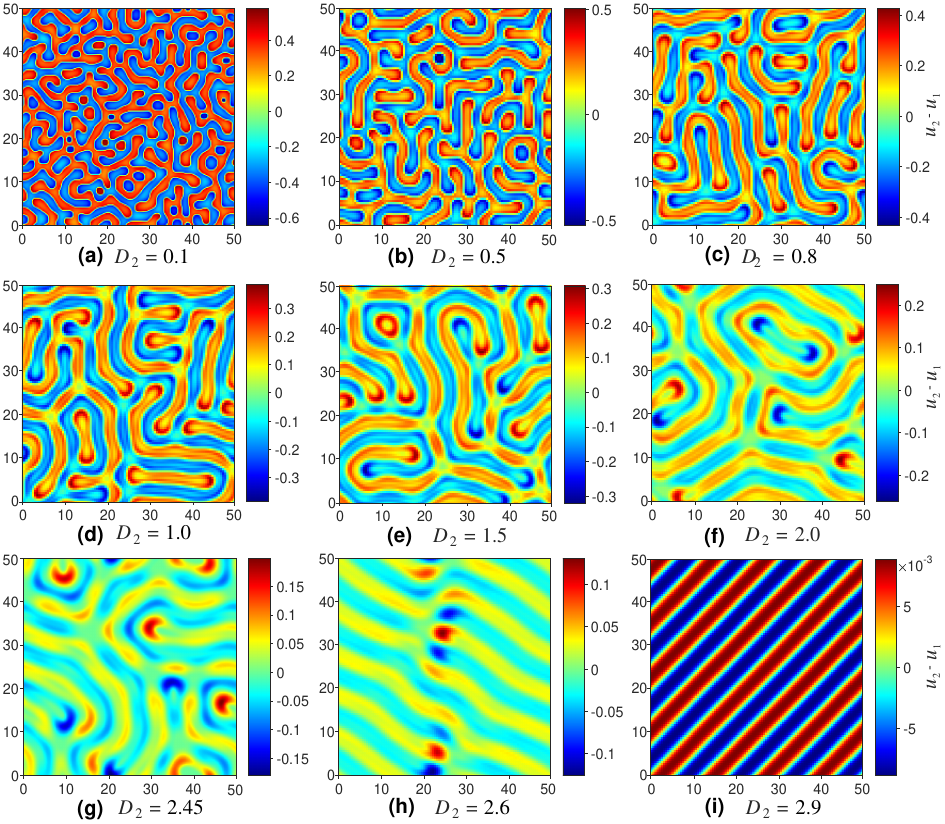}
	\caption{Interlayer difference \(\Delta u=u_2-u_1\) for the stripe states. Same \(D_2\) values as in Fig.~\ref{fig:stri_u2}. The amplitude of \(\Delta u\) decreases monotonically with increasing \(D_2\), and the color scale shrinks by an order of magnitude from \(D_2=0.1\) to \(D_2=2.9\). This trend reflects the narrowing of the Turing-unstable band in Eq.~\eqref{eq:char} and the associated enhancement of interlayer synchrony, which is the dynamical counterpart of the pathological synchronization targeted by desynchronizing DBS~\cite{tass2003}.}
	\label{fig:stri_difference}
\end{figure}
	\clearpage
\begin{figure}[ht!]
	\centering
	\includegraphics[width=\columnwidth]{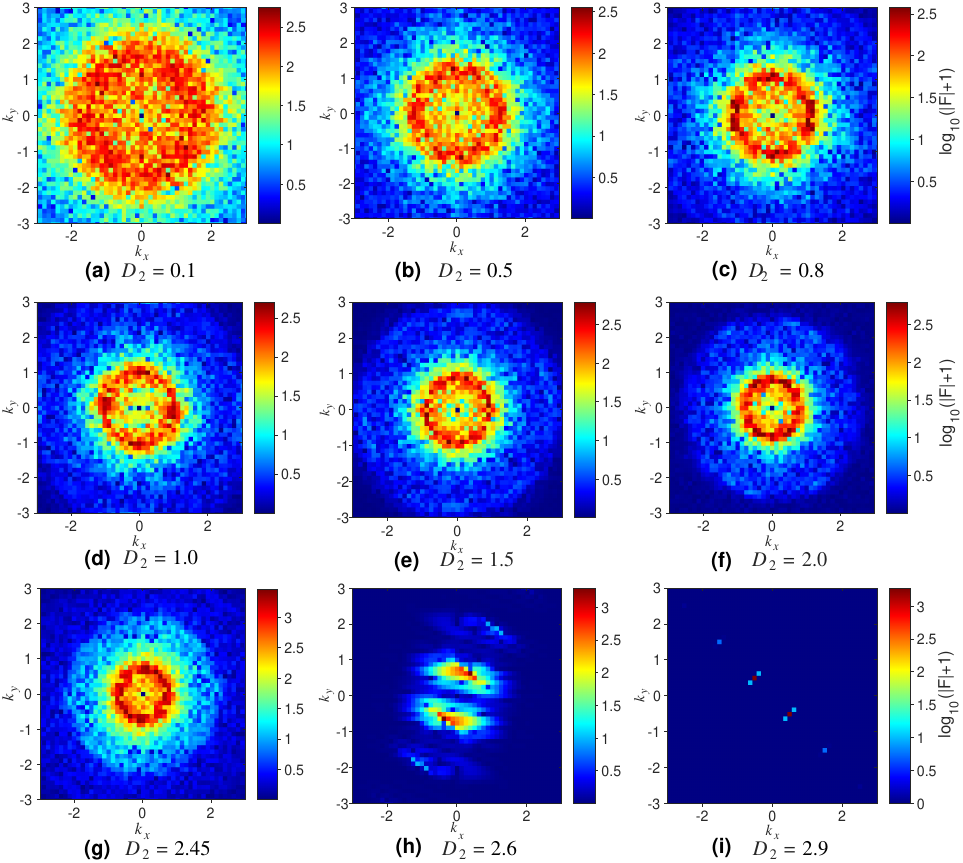}
	\caption{Spatial power spectrum \(\log_{10}(|F|+1)\) of the stripe states for the same \(D_2\) values. For \(D_2\lesssim2\), the spectrum is a broad ring centered at the most unstable wavenumber. As \(D_2\to3^-\), the ring contracts into discrete peaks, reflecting the narrowing of the unstable band and the selection of a single dominant wavelength. The peak wavenumber is consistent with the linear prediction \(k_c(D_2)\) from Eq.~\eqref{eq:kc}; the sharpness of the peaks quantifies the degree of spatial order of the Turing pattern~\cite{cross1993}.}
	\label{fig:stri_power}
\end{figure}
	\clearpage
\begin{figure}[ht!]
	\centering
	\includegraphics[width=\columnwidth]{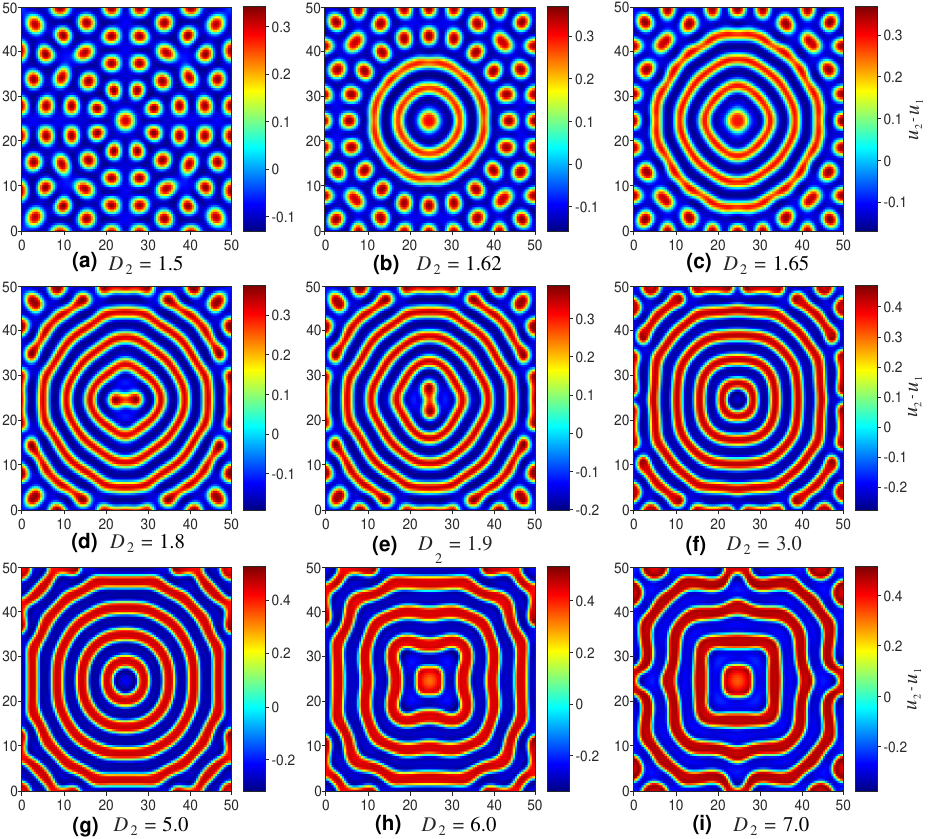}
	\caption{Interlayer difference \(\Delta u=u_2-u_1\) for the spot-to-ring states shown in Fig.~\ref{fig:dot_circle_u1}. Same \(D_2\) values as in Fig.~\ref{fig:dot_circle_u1}. Throughout the spot-to-ring transition, the amplitude of \(\Delta u\) remains confined to a narrow range and does not exhibit an order-of-magnitude change, in contrast to the monotonic decrease observed for stripes (see Fig.~\ref{fig:stri_difference}) and the monotonic increase observed for radial modes (see Fig.~\ref{fig:babo_difference}). 
	This indicates that the spot-to-ring transition is primarily a rearrangement of the spatial structure rather than a change in the degree of interlayer synchrony.}
	\label{fig:dot_circle_difference}
\end{figure}
	\clearpage
\begin{figure}[ht]
	\centering
	\includegraphics[width=\columnwidth]{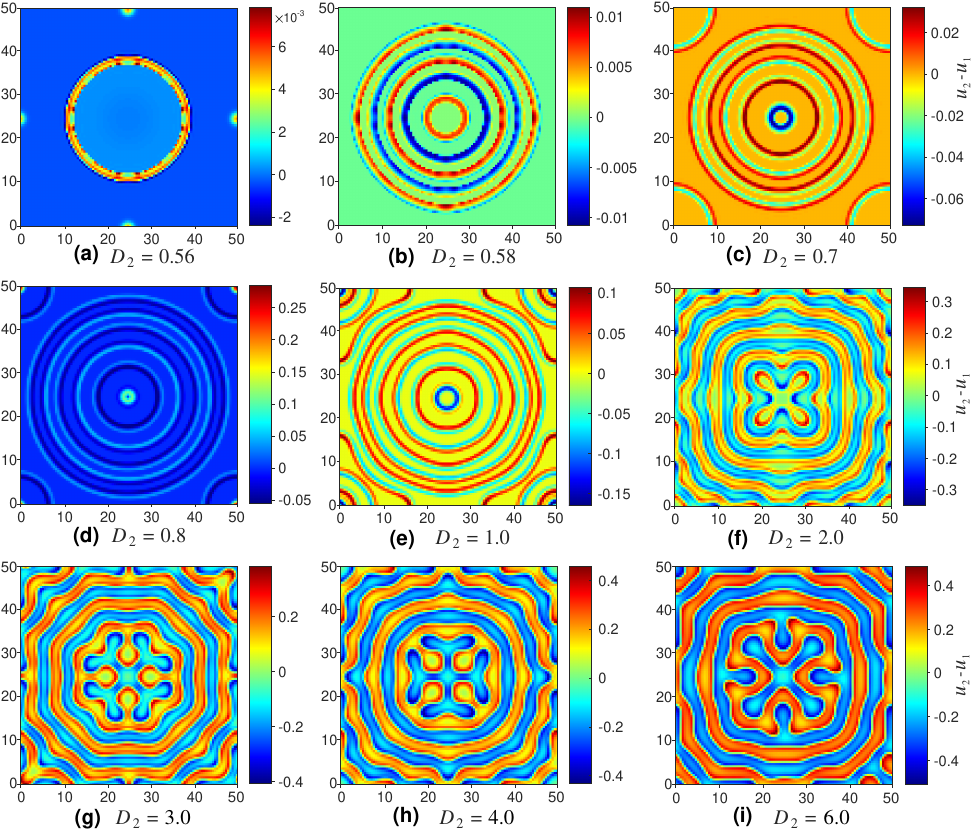}
	\caption{Interlayer difference \(\Delta u=u_2-u_1\) for the radial modes shown in Fig.~\ref{fig:babo_u1}. 
		Same \(D_2\) values as in Fig.~\ref{fig:babo_u1}. 
		The amplitude of \(\Delta u\) increases by more than two orders of magnitude, from \(\sim10^{-3}\) at \(D_2=0.56\) to \(\sim0.4\) at \(D_2=6.0\). 
		This growth indicates that increasing \(D_2\) progressively weakens interlayer synchrony in the radial regime, in contrast to the stripe regime where increasing \(D_2\) enhances synchrony. 
		The opposite trends suggest that the therapeutic effect of diffusion-like modulation depends qualitatively on the selected pattern morphology.}
	\label{fig:babo_difference}
\end{figure}

	\clearpage
\begin{figure}[ht!]
	\centering
	\includegraphics[width=16cm]{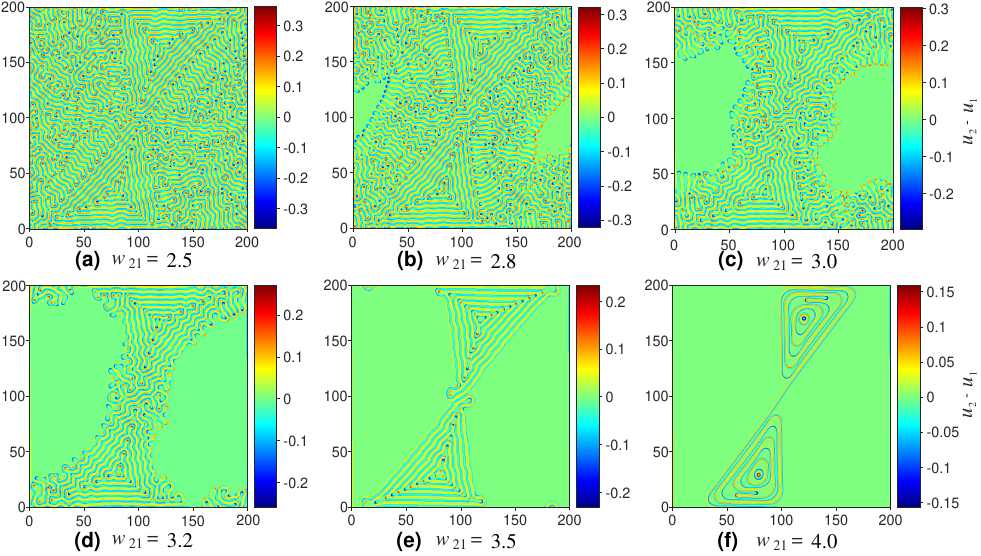}
	\caption{Interlayer difference \(\Delta u=u_2-u_1\) for the spiral-wave states shown in Fig.~\ref{fig:Spiral_u1}. Same \(w_{21}\) values as in Fig.~\ref{fig:Spiral_u1}. Across the defect-mediated turbulent regime [\(w_{21}\le3.0\)] and the isolated-spiral regime [\(w_{21}\ge3.5\)], the amplitude of \(\Delta u\) decreases only by a factor of two, from \(\sim0.3\) to \(\sim0.15\), in contrast to the order-of-magnitude changes observed in the stripe and radial regimes (see Figs.~\ref{fig:stri_difference} and~\ref{fig:babo_difference}). The weak dependence of \(\Delta u\) on \(w_{21}\) indicates that the transition from turbulence to isolated spirals is primarily a change in the number and arrangement of defects, rather than a change in the degree of interlayer synchrony.}
	\label{fig:Spiral_difference}
\end{figure}

\begin{figure}[ht!]
	\centering
	\includegraphics[width=16cmh]{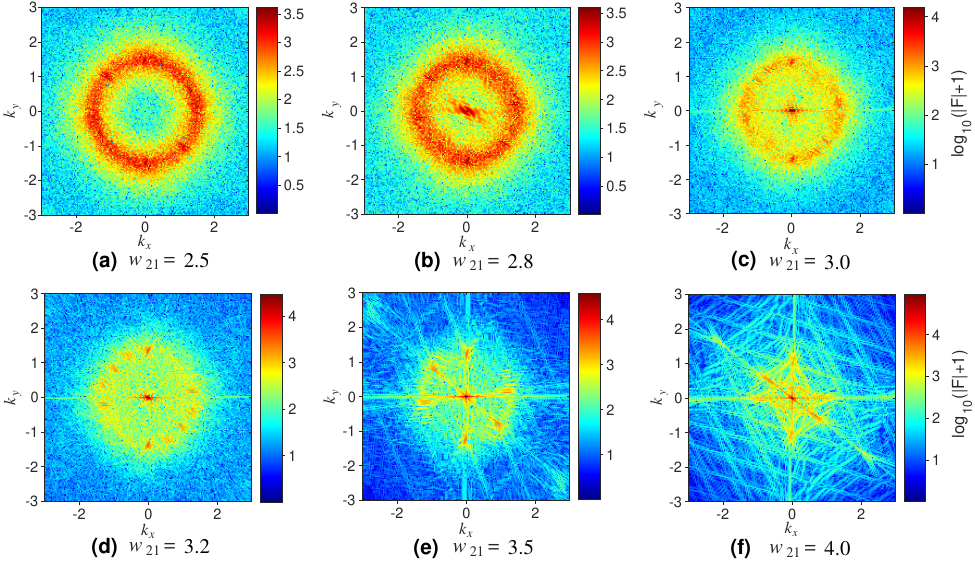}
	\caption{Spatial power spectrum \(\log_{10}(|F|+1)\) of the spiral-wave states for the same \(w_{21}\) values as in Fig.~\ref{fig:Spiral_u1}. For \(w_{21}\le3.0\), the spectrum is a broad ring centered at a finite wavenumber, characteristic of defect-mediated turbulence. As \(w_{21}\) increases, the ring contracts into discrete peaks and, for \(w_{21}=3.5\) and \(4.0\), into a small set of dominant modes with additional streak-like features arising from the symmetry of the domain. The spectral narrowing reflects the reduction of the number of active modes and the emergence of well-separated, isolated spiral cores.}
	\label{fig:Spiral_power}
\end{figure}

	\clearpage
\begin{figure}[t]
	\centering
	\includegraphics[width=\columnwidth]{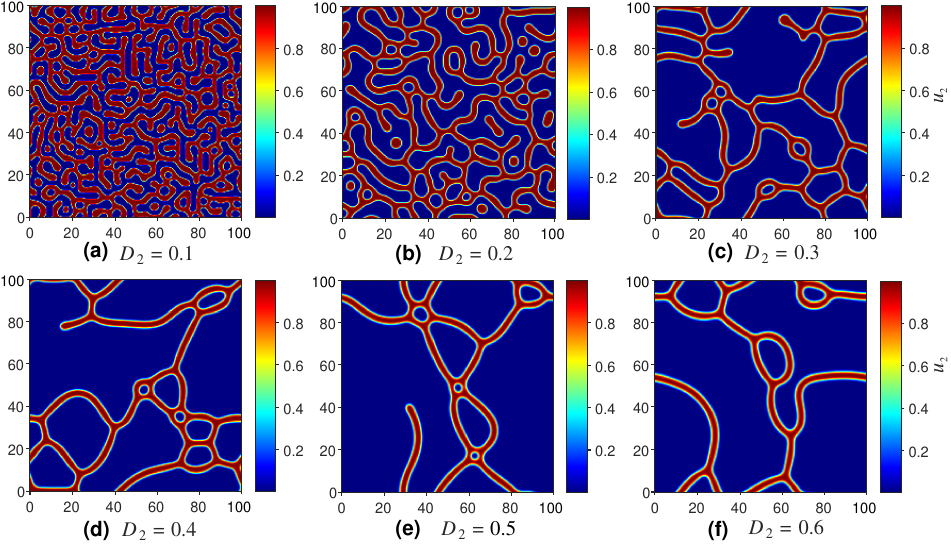}
	\caption{Branched, neuron-like morphologies in the lower layer \(u_2(\mathbf{r},t)\) for \(D_2=0.1,0.2,0.3,0.4,0.5,0.6\). 
		The morphology closely mirrors that of the upper layer (see Fig.~\ref{fig:neural_u1}), and the dendrite-like branches coarsen and merge as \(D_2\) increases. 
		The upper and lower layers thus share the same spatial organisation across the whole range of \(D_2\).}
	\label{fig:neural_u2}
\end{figure}

\begin{figure}[t]
	\centering
	\includegraphics[width=\columnwidth]{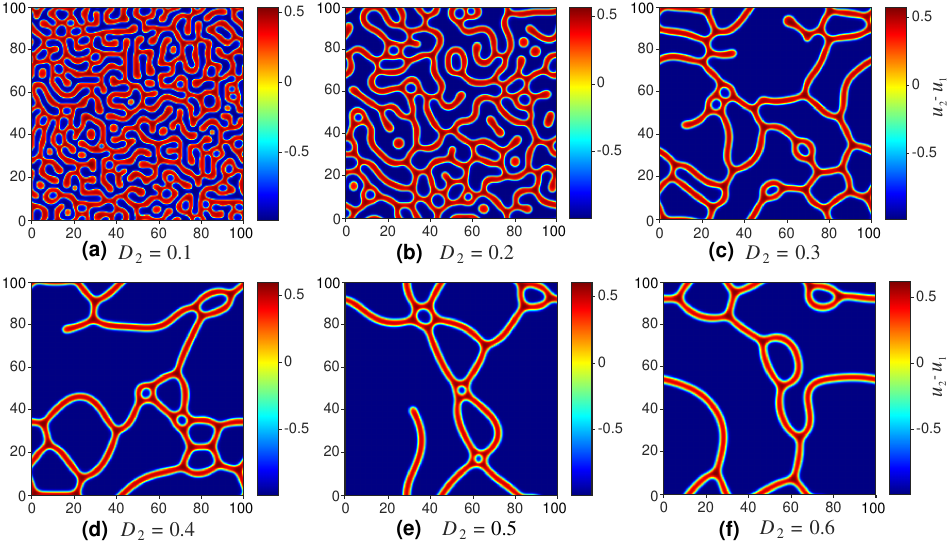}
	\caption{Interlayer difference \(\Delta u=u_2-u_1\) for the neuron-like states shown in Figs.~\ref{fig:neural_u1} and~\ref{fig:neural_u2}. 
		Same \(D_2\) values as in Fig.~\ref{fig:neural_u2}. 
		Throughout the coarsening of the branched morphology, the amplitude of \(\Delta u\) remains confined to the range \(0\lesssim\Delta u\lesssim0.5\), with no order-of-magnitude change. 
		This behaviour is intermediate between the stripe and radial regimes: the morphology reorganises as \(D_2\) grows, but the degree of interlayer synchrony is essentially conserved.}
	\label{fig:neural_difference}
\end{figure}

%\begin{figure}[ht!]
%	\centering
%	\includegraphics[width=\columnwidth]{neural_power.pdf}
%	\caption{  }
%	\label{fig:neural_power}
%\end{figure}

\end{document}